\pdfoutput=1
\RequirePackage{ifpdf}
\ifpdf 
\documentclass[pdftex]{sigma}
\else
\documentclass{sigma}
\fi

\numberwithin{equation}{section}

\usepackage{verbatim,bm,mathtools,empheq,euscript}

\newcommand{\rmd}{\mathrm{d}}
\newcommand{\dee}{\mathrm{d}}
\newcommand{\A}{\EuScript{A}}

\newcommand{\D}{\EuScript{D}}

\newcommand{\R}{\EuScript{R}}
\newcommand{\T}{\EuScript{T}}
\newcommand{\U}{\EuScript{U}}
\newcommand{\V}{\EuScript{V}}

\DeclareMathOperator{\Ber}{Ber}

\newcommand{\wt}{\widetilde}
\newcommand{\h}{\widehat}

\newcommand{\q}{\mathrm{q}}

\newcommand{\suminnt}{\sum \hspace{-3ex}\int}

\newcommand{\dfn}{\vcentcolon=}

\DeclarePairedDelimiterX{\av}[1]{\langle}{\rangle}{#1}

\usepackage{lmodern}

\DeclareSymbolFont{euletters}{U}{zeur}{m}{n}
\DeclareMathSymbol{\g}{\mathalpha}{euletters}{`\g}
\DeclareMathSymbol{\n}{\mathalpha}{euletters}{`\n}
\DeclareMathSymbol{\m}{\mathalpha}{euletters}{`\m}
\DeclareMathSymbol{\B}{\mathalpha}{euletters}{`\b}

\begin{document}

\allowdisplaybreaks

\newcommand{\arXivNumber}{2307.06355}

\renewcommand{\PaperNumber}{090}

\FirstPageHeading

\ShortArticleName{Moving NS Punctures on Super Spheres}

\ArticleName{Moving NS Punctures on Super Spheres}

\Author{Dimitri P.~SKLIROS}

\AuthorNameForHeading{D.P.~Skliros}

\Address{Blackett Laboratory, Imperial College London, SW7 2AZ, UK\footnote{Visitor status.}}
\Email{\href{mailto:dp.skliros@gmail.com}{dp.skliros@gmail.com}}

\ArticleDates{Received February 20, 2024, in final form September 18, 2024; Published online October 10, 2024}

\Abstract{One of the subtleties that has made superstring perturbation theory intricate at high string loop order is the fact that as shown by Donagi and Witten, supermoduli space is not holomorphically projected, nor is it holomorphically split. In recent years, Sen (further refined by Sen and Witten) has introduced the notion of vertical integration in moduli space. This enables one to build BRST-invariant and well-defined amplitudes by adding certain correction terms to the contributions associated to the traditional ``delta function'' gauge fixing for the worldsheet gravitino on local patches. The Sen and Witten approach is made possible due to there being no obstruction to a smooth splitting of supermoduli space, but it may not necessarily be the most convenient or natural solution to the problem. In particular, this approach does not determine what these corrections terms actually are from the outset. Instead, it shows that such correction terms in principle exist, and when included make all perturbative amplitudes well-defined. There may be situations however where one would like to instead have a well-defined and fully determined path integral at arbitrary string loop order from the outset. In this paper, I initiate an alternative (differential-geometric) approach that implements the fact that a smooth gauge slice for supermoduli space always exists. As a warmup, I focus specifically on super Riemann surfaces with the topology of a sphere in heterotic string theory, incorporating the corresponding super curvature locally, and introduce a new well-defined smooth gauge fixing that leads to a globally defined path integral measure that translates arbitrary fixed ($-1$) picture NS vertex operators (or handle operators) (that may or may not be offshell) to integrated (0) picture. I also provide some comments on the extension to arbitrary super Riemann surfaces.}


\Classification{32G05; 32G15; 51M15; 53Z05}

\section{Introduction}

One of the most basic starting points for superstring perturbation theory is the notion of a~vibrating loop of string, suitably formulated so as to naturally incorporate the elementary principles of quantum mechanics and relativity. A loop of string, in turn, has left- and right-moving degrees of freedom which turn out to be largely independent \cite{Polchinski_v1,Witten12c}. Famously, in the heterotic string~\mbox{\cite{GrossHarveyRohm85a,GrossHarveyRohm85b}} this distinction between chiral and anti-chiral halves is so stark that (in one formulation) we can even regard the two halves as living in a different number of spacetime dimensions \cite{Polchinski_v2,Witten12c}. At tree level, this ``chiral factorisation'' leads to a whole host of interesting developments, such as the celebrated Kawai--Lewellen--Tye (KLT) relations \cite{KLT86}, which have in turn led to a whole range of progress (see \cite{BernCarrascoChiodaroliJohanssonRoiban19} and references therein), such as BCJ duality, double copy constructions, colour-kinematics dualities, ambitwistor strings, and various other recent incarnations; see also \cite{Mizera17} and references therein. There is also recent work on a one-loop version of the KLT relations \cite{Stieberger22} and corresponding double copy structure \cite{Stieberger23}. Although this has also led to vast simplifications, enabling, e.g., one to carry out computations at high loop orders in the context of supergravity, the string theory understanding thereof at high loop order is much less understood or developed.

In the full string theory context, at loop level, a chiral factorisation (at least at the level of integrands) that is reminiscent of the KLT relations is also present under certain assumptions. This in turn led to the D'Hoker and Phong \emph{chiral splitting} theorem \cite{D'HokerPhong89} (elaborated on in detail and for general constant backgrounds in bosonic string theory in \cite{SklirosCopelandSaffin17}). This theorem is used in a~number of recent applications \cite{DHokerMafraPiolineSchlotterer20,GeyerMonteiroStark-Much21}, including the celebrated 2-loop calculations \cite{D'HokerPhongI}. Briefly, this theorem states that (super)string integrands chirally factorise when we hold the loop momenta (and Dirac zero modes when present) fixed. It however relies on the Belavin--Knizhnik theorem \cite{BelavinKnizhnik86} which is a statement about the chiral factorisation of the ghost or superghost contributions to the superstring path integral measure when the total central charge vanishes. And this in turn is based on the assumption that supermoduli space is holomorphically split, an assumption that has been suspected to be incorrect for decades \cite{AtickMooreSen88,Nelson88b,VerlindeHphd}, and is now known to break down at sufficiently high genus as shown by Donagi and Witten in~\cite{DonagiWitten15}.\looseness=1

So it turns out that superstring amplitudes, in particular, do not chirally factorise beyond a sufficiently high number of string loops. The associated global obstruction is due to~\cite{DonagiWitten15} supermoduli space not being holomorphically split or holomorphically projected (see also~\cite{VerlindeHphd} and especially~\cite{Nelson88b} for some early work on this). To elaborate a little, in the RNS approach~\cite{Polchinski_v2,VerlindeVerlinde87b,VerlindeHphd,Witten12b,Witten12a,Witten12c} to superstring perturbation theory one usually begins by considering embeddings of super Riemann surfaces of a fixed genus into spacetime (and/or a more abstract target space associated to the superconformal field theory of interest). After integrating over all such embeddings, we are to integrate over the corresponding supermoduli space, before finally summing over string loops \cite{Polchinski_v2,Witten12c}. To every point in supermoduli space there is a corresponding equivalence class of super Riemann surfaces (super Riemann surfaces related by a superconformal transformation are deemed equivalent). Briefly, the problem is associated to the non-vanishing of certain cohomology classes \cite{Nelson88b}. In practice, this manifests as follows. Supermoduli space $\mathfrak{M}$ in general requires several coordinate charts \smash{$\big(\U_m,\wt{\tau}^{\tilde{\ell}}_m;\tau_m^\ell | \chi_m^\alpha\big)$} (with a~collection of open sets $\{\U_m\}$ covering $\mathfrak{M}$, and \smash{$\wt{\tau}_m^{\tilde{\ell}}$}, \smash{$\tau_m^\ell$} corresponding to even moduli and $\chi_m^\alpha$ odd local coordinates), and it is not possible in general to find holomorphic transition functions on patch overlaps $\U_m\cap \U_n$ that preserve the $\mathbb{Z}$ grading \cite{Nelson88b} as in
\begin{gather*}
\wt{\tau}_m^{\tilde{\ell}} = \wt{f}_{mn}^{\tilde{\ell}}\big(\wt{\tau}_n^1,\wt{\tau}_n^2,\dots\big),\\
\tau^\ell_m=f_{mn}^\ell\big(\tau^1_n,\tau_n^2,\dots\big), \qquad \qquad \text{(holomorphic splitting)}\\
\chi^\alpha_m=\sum_\beta \chi^\beta_nh^{\alpha\beta}_{mn}\big(\tau^1_n,\tau^2_n,\dots\big),
\end{gather*}
which would correspond to having found a \emph{holomorphic splitting}, which in turn does not exist in general. The Grassmann-even quantities $\wt{f}_{mn}$, $f_{mn}$ and $h_{mn}$ are transition functions defined on patch overlaps $\U_m\cap \U_n$ depending on the even coordinates $\wt{\tau}^{\tilde{\ell}}_n$, $\tau_n^\ell$ of the $\big(\U_n,\wt{\tau}^{\tilde{\ell}}_n;\tau_n^\ell| \chi_n^\alpha\big)$ chart as indicated. It is also not possible to find an atlas that preserves the $\mathbb{Z}$ grading of the even coordinates only, as in
\begin{gather*}
\wt{\tau}_m^{\tilde{\ell}} = \wt{f}_{mn}^{\tilde{\ell}}\big(\wt{\tau}_n^1,\wt{\tau}_n^2,\dots\big),\\
\tau^\ell_m=f_{mn}^\ell\big(\tau^1_n,\tau_n^2,\dots\big),\qquad\qquad\qquad \text{(holomorphic projection)}\\
\chi^\alpha_m= g^\alpha_{mn}\big(\tau^1_n,\tau^2_n,\ldots| \chi^1_n,\chi_n^2,\dots\big),
\end{gather*}
where the odd transition function $g^\alpha_{mn}$ only preserves the $\mathbb{Z}$ grading of the odd supermoduli~$\chi_m^\alpha$ mod~2. This corresponds to having found a holomorphic projection, which is also known to not exist in general \cite{DonagiWitten15}. What does exist instead is an atlas whose charts are glued together with both even and odd transition functions that only preserve the $\mathbb{Z}$ grading mod 2
\begin{gather}
\wt{\tau}_m^{\tilde{\ell}} = \wt{f}_{mn}^{\tilde{\ell}}\big(\wt{\tau}_n^1,\wt{\tau}_n^2,\dots;\tau^1_n,\tau^2_n,\ldots| \chi^1_n,\chi_n^2,\dots\big),\nonumber\\
\tau^\ell_m=f_{mn}^\ell\big(\wt{\tau}_n^1,\wt{\tau}_n^2,\dots;\tau^1_n,\tau^2_n,\ldots| \chi^1_n,\chi_n^2,\dots\big),\qquad \text{(general)}\nonumber\\
\chi^\alpha_m= g^\alpha_{mn}\big(\wt{\tau}_n^1,\wt{\tau}_n^2,\dots;\tau^1_n,\tau^2_n,\ldots| \chi^1_n,\chi_n^2,\dots\big),
\label{eq:generaltransfunc}
\end{gather}
so that $\wt{f}_{mn}^{\tilde{\ell}}$, $f_{mn}^\ell$ are even Grassmann parity smooth functions of their arguments, whereas~$g^\alpha_{mn}$ are odd parity smooth functions of their arguments. The range of the various indices ${\wt{\ell}\!=\!1,\dots,\wt{\m}}$, $\ell=1,\dots,\m$ and $\alpha=1,\dots,\nu$, correspond to the relevant dimension, ${\rm even}|{\rm odd}=\wt{\m}+\m|\nu$, of supermoduli space, whereas $m$, $n$ label the charts. (In all cases there are also the obvious compatibility requirements or cocycle relations associated to triple and higher patch overlaps.) In other words, there is no obstruction\footnote{I thank Edward Witten for some correspondence on this.} to a \emph{smooth splitting} of supermoduli space. In fact, \emph{all} obstructions to a splitting vanish on a smooth supermanifold \cite{Nelson88b}, in that we can always interpolate between one sort of behaviour near the boundary of a patch and another on the interior \cite{Nelson88b}.\looseness=-1

There is some very important work that implements this observation, initiated by Sen \cite{Sen15b} and further refined by Sen and Witten \cite{SenWitten15} (see \cite[Section~7.1]{SenWitten15}), see also Wang and Yin in \cite{WangYin22}. There is also a related more algebraic approach carried out by Erler in \cite{ErlerKonopka17}. These developments are important because we did not have an explicit prescription in the RNS formalism to compute higher loop superstring amplitudes prior to it. (Although one should be careful, because this does not mean that vertical integration is necessarily the best way to proceed, nor that it is particularly natural.\footnote{To clarify (jumping ahead slightly), I am using the word ``natural'' here to refer to the fact that (given the above discussion associated to smooth and split transition functions), even and odd moduli naturally mix as we transition from one patch to another, whereas in the PCO formalism one has implicitly integrated out the odd moduli from the outset, making a geometric interpretation unclear. In the differential-geometric formalism of the current document, one instead treats even and odd moduli on equal footing, so that one can transition from one patch to a neighbouring one using smooth transition functions of the form \eqref{eq:generaltransfunc}. (We also treat even and odd coordinates of the underlying super Riemann surface on equal footing, instead of reducing to an ordinary Riemann surface from the outset as done in the PCO approach.) So one might consider this a more ``natural'' procedure, but of course ultimately the PCO and differential-geometric approaches should lead to the same observable physics.}) The idea \cite{Sen15b,SenWitten15} is to work in the picture-changing operator (PCO) formalism, where one has picked a local (e.g., delta function) gauge slice for the worldsheet gravitino that may or may not {\it a priori} be globally well-defined. The odd moduli are then integrated out, leading to PCO insertions on an ordinary Riemann surface which is partitioned into regions. The locations of PCOs are chosen such that spurious poles (associated to the incorrect gauge fixing) in each region are avoided. After carrying out this procedure for every region, if one tries to simply add the contributions from every region one finds that the resulting quantity is not well-defined. This manifests in a number of ways, e.g., amplitudes are not gauge invariant (BRST-exact vertex operators do not decouple)~\cite{Sen15b}.
The prescription, dubbed ``vertical integration'' \cite{Sen15b,SenWitten15}, is to nevertheless go ahead and add the contributions from each region, and then to correct for the incorrect gauge fixing by including correction terms associated to the interface between regions. This effectively connects the aforementioned locally-defined sections along a fibre (corresponding to a coordinate choice) over a fixed point in moduli space. In this manner gauge invariance is restored. Vertical integration, effectively, makes use of the fact that a smooth splitting of supermoduli space always exists. But the situation is not entirely satisfactory yet \cite{BerkovitsDHokerGreenJohanssonSchlotterer22}. For example, the vertical integration procedure does not explicitly determine what these correction terms actually are.

The question I would like to ask in this note therefore is how to construct a smooth gauge slice for the integral over supermoduli in the context of heterotic string theory that is determined and well-defined from the outset. In fact, I will consider the perhaps simplest example of such a smooth gauge slice, namely I will derive how to translate a Neveu--Schwarz (NS) puncture across a super Riemann surface with the topology of a 2-sphere, with special attention paid to incorporating the underlying super curvature \emph{ locally}. The gauge slice will be defined by making use of a super Riemann surface analogue of a metric, and I will choose a specific superconformal frame that is globally well-defined. In other words, I will introduce a metric on [a subset of] supermoduli space and (after a suitable gauge fixing) use it to induce a smooth dependence of superconformal transition functions (defining the super Riemann surface) on the supermoduli associated to the location of an NS puncture. The gauge fixing will be analogous to Polchinski's ``as flat as possible'' gauge slice in the bosonic string theory context~\cite{Polchinski88}~-- reviewed in detail in~\cite{SklirosLuest21}~-- suitably generalised to heterotic super Riemann surfaces.

So in the current note I initiate a \emph{differential-geometric} approach to supermoduli space, that is well-defined from the outset, that does not rely on existence of a holomorphic splitting or projection. A central role will be played by the local super curvature, which in a sense localises the ``Wu--Yang'' type contributions from patch overlaps, so that total derivatives in supermoduli space really do correspond to total derivatives in supermoduli space (as opposed to integrals of total derivatives that receive contributions from fictitious boundaries associated to patch overlaps \cite{VerlindeHphd}). (The mechanism is analogous to a baby version of the integral reviewed in \cite[Section~5.6]{SklirosLuest21} in the context of bosonic string theory.) Note that because we are allowing for non-trivial super curvature, the point at infinity on the super plane is in a sense trivialised. So that (in a practical calculation) we only really need one coordinate chart for the entire sphere. This is to be contrasted with the approach of Nelson \cite{Nelson89} and La and Nelson \cite{LaNelson90}, and hence, more generally, also (super)string field theory \cite{deLacroixErbinKashyapSenVerma17} which incorporates Nelson's viewpoint; see also Mark Doyle's thesis\footnote{I am grateful to Mark Doyle for sharing his Ph.D.\ Thesis with me.} \cite{DoylePhD} for a very nice overview of this approach. In this approach \cite{Nelson89}, local super curvature is instead hidden in transition functions on patch overlaps. For example, in the case of a super sphere it might be encoded on an equatorial band, making it essential to consider more than a single chart.

Although smoothly translating an NS puncture across a super sphere is perhaps the simplest non-trivial example, this could nevertheless be expected to have wide-ranging implications. Because in addition to providing the relevant map from fixed picture $-1$ NS vertex operators to integrated 0 picture, this also (partially) provides an explicit expression for the measure associated to translating heterotic string \emph{handle operators} across a curved surface. Modulo the inclusion of the Ramond sector (that I will not discuss here), the latter provides the basic building block of arbitrarily higher-genus superstring amplitudes, and (due to the underlying smooth gauge slice \cite{Nelson88b}) there will be no obstruction at any string loop order. So that amplitudes at all loop orders can be treated on equal footing. As I briefly \emph{speculate} on next, it is not inconceivable that this approach therefore also proves useful beyond perturbation theory.\looseness=-1

With regard to this last point, one cannot dismiss from the outset the possibility that this differential-geometric handle operator approach might even make aspects of non-perturbative string theory accessible. It is of course very well-known that superstring perturbation theory is an asymptotic series \cite{Shenker90}. And although there is no proof, one expects that string perturbation theory diverges for the same reason that almost any Feynman diagram expansion diverges, and one simply does not expect there to be any surprises in the string case.\footnote{I am grateful to Edward Witten for a discussion about this over dinner at STRINGS 2014 in Princeton.} That is, if one were able to compute the genus-$\g$ contribution to the amplitude, for every $\g=0,1,2,\dots$, then the resulting series should be asymptotic. The genus-$\g$ contribution to the amplitude with $\n$ vertex operator insertions might in particular be expected to diverge as $\mathcal{A}_{\g,\n}\sim (2\g-3+\n)!$. The underlying reasoning leading to this conclusion however does not apply in the handle operator viewpoint, because in the handle operator viewpoint I would like to suggest that one instead might like to sum over loops (handle operator insertions) \emph{before} carrying out the path integral, i.e., at the level of the integrand. And there is no reason to believe that the procedure of summing over string loops, and that of carrying out the path integral, commute. In further detail, by adopting an appropriate unfolding\footnote{This would be analogous to the unfolding procedure carried out by Mirzhakani \cite{Mirzakhani06,Mirzakhani07}, see also \cite{DijkgraafWitten18}, in the context of hyperbolic Riemann surfaces, and the related procedure explained in \cite{O'BrienTan87} and \cite{AngelantonjFlorakisPioline12}; see also citations `to' and references `within' \cite{AngelantonjFlorakisPioline12}. None of these procedures applies directly however, because the handle operator gauge slice is formulated on (super) Riemann surfaces with the topology of a 2-sphere.} of supermoduli space and rewriting it as an integral over super Teichm\"uller space (so as to remove the correlations between the integration domains of the various handle operator supermoduli integrals), one might try to first sum over handle operator insertions before carrying out the string path integral. A simple toy model analogy, to get the idea across, would be to consider the integral of a Gaussian with a $x^4$ interaction term on the right-hand side of the equality
\begin{gather}\label{eq:SumInt!=IntSum}
\sum_{\g=0}^\infty \frac{1}{\g!}\int_{-\infty}^\infty {\rm d}x\, {\rm e}^{-xGx}\bigl(-\lambda x^4\bigr)^{\g} \neq
\int_{-\infty}^\infty {\rm d}x \, {\rm e}^{-xGx}\sum_{\g=0}^\infty \frac{1}{\g!}\bigl(-\lambda x^4\bigr)^{\g}.
\end{gather}
Although the left-hand side of this expression is an asymptotic series, the sum over $\g$ on the right-hand side is not: both the sum and corresponding integrals can be carried out without encountering a divergence associated to the presence of an asymptotic series; in fact, the right-hand side can be written in terms of a modified Bessel function of the second kind, \smash{$ \frac{1}{2}\sqrt{\frac{G}{\lambda}}{\rm e}^{\frac{G^2}{8\lambda}}K_{\frac{1}{4}}\big(\frac{G^2}{8\lambda}\big)$}. The handle operator approach enables one to explore the idea that a correct approach might be to think of string theory as being defined by an expression analogous to the right-hand side in~\eqref{eq:SumInt!=IntSum}, instead of the usual approach which corresponds to integrating the summand at fixed $\g$ and then trying to sum the asymptotic series (as seen on the left-hand side). One might argue that this idea was already hinted at by the Fischler--Susskind mechanism \cite{FischlerSusskind86a, FischlerSusskind86b, Polchinski88}, which highlights a~subtle interplay between string loop corrections and string background shifts, and a~little-known paper by Tseytlin \cite{Tseytlin90}, but the details have not been fleshed out. Of course, there are also IR divergences arising from the boundary of supermoduli space, where one understands in principle how to proceed due to developments in string field theory, but it is not yet clear how these divergences will be dealt with in the handle operator viewpoint. There is clearly a~lot of work to be done.

In Section~\ref{sec:SRS}, we begin by discussing a simple parametrisation of a 2-sphere using both a~holomorphic and a smooth viewpoint. In Section~\ref{sec:SCNC}, we define a new set of frame coordinates that we call `superconformal normal coordinates'. These coordinates (by analogy to Riemann normal coordinates in ordinary Riemannian geometry) will enable us to map (or pullback) our superframe (on which we use standard superconformal field theory techniques to define local operators and states) to an underlying curved super Riemann surface. In Section~\ref{sec:SC}, we introduce the notion of super curvature that we will adopt throughout the article. In Section~\ref{sec:PIM}, we will derive the precise expression for the path integral measure that implements the aforementioned gauge slice. In Section~\ref{sec:GI}, we show that the path integral measure contributions associated to the gauge slice of interest leads to the expected decoupling of BRST-exact insertions into the path integral. In the Discussion (see Section~\ref{sec:D}), we provide some further context and generalisation to arbitrary super Riemann surfaces with an arbitrary number of handles while also highlighting a puzzle that seems to arise in this case. We also mention some future directions.

\section{Super Riemann surfaces}\label{sec:SRS}

Let us primarily construct a super Riemann surface, $\Sigma$, with the topology of a 2-sphere. We can, for instance, glue two copies of the super plane (see also \cite[Section~5.2.1]{Witten12a}) that are in turn parametrised by the superconformal charts $(\U_u,u|\theta_u)$ and $(\U_w,w|\psi)$. We take these to be centred at $\q_u\in S^2$ and $\q_w\in S^2$ respectively so that $(u|\theta_u)(\q_u)=0$ and $(w|\psi)(\q_w)=0$. We might think of these points as corresponding to north and south poles of the reduced 2-sphere. We then glue on $\U_u\cap \U_w$ (which loosely speaking can be thought of as spanning an equatorial band) with the superconformal transition function, $uw=1$. Demanding consistency with the superconformal condition $D_{\psi}u=\theta_u D_{\psi}\theta_u$ in turn determines the remaining transition function $\theta_u(w|\psi)$ (up to an immaterial sign since both the superconformal condition and $uw=1$ are invariant under $u|\theta_u\rightarrow u|-\theta_u$ and $w|\psi\rightarrow w|-\psi$). Proceeding in a similar manner for the anti-chiral half, the full set of transition functions to cover the entire sphere is then \cite{Polchinski_v2,Witten12b},
\begin{gather}\label{eq:ztheta-S2-trans}
\wt{u}(\wt{w}) =\frac{1}{\wt{w}},\qquad
u(w|\psi) = \frac{1}{w}
\qquad\text{and}\qquad
\theta_u(w|\psi)= {\rm i}\frac{\psi}{w},
\end{gather}
where we arbitrarily picked one of the two signs (the alternative choice is effectively equivalent to replacing the explicit factor, ${\rm i}\rightarrow 1/{\rm i}$). In solving the superconformal condition one makes use of the fact that it holds for all $\psi$.

To transition to a smooth description, it is convenient to use the above charts, $(\U_u,u|\theta_u)$ and~${(\U_w,w|\psi)}$, to construct a super Riemann surface version of a metric on the sphere. The super analogue of a Riemannian metric has been provided in \cite[Section~3.6.3]{Witten12b}, see also \cite{BaranovFrolovShvarts87,VerlindeHphd} for some early work along these lines. As discussed in \cite{Witten12b} (and elaborated on very briefly in Section~\ref{sec:SC} here where we also introduce a notion of super curvature), the appropriate structure, locally (where we pick local coordinates $\wt{z};\!z|\theta$) is the following. We can regard a metric as a nonzero section $\wt{E}={\rm e}^{\tilde{\varphi}}\rmd\wt{z}$ of $T^*_L\Sigma$ (of rank $1|0$), and a nonzero section $E={\rm e}^{\varphi}\varpi$ (with~${\varpi = \rmd z-\rmd \theta\theta}$) of a~rank~$1|0$ subbundle, $\mathcal{D}^{-2}\subset T^*_R\Sigma$ (with~${T^*_R\Sigma}$ of rank $1|1$). We can then introduce a~connection~$\omega$ on $\mathcal{D}^{-2}$, and a corresponding gauge invariance $E\rightarrow {\rm e}^uE$, $\wt{E}\rightarrow {\rm e}^{-u}\wt{E}$ and $\omega\rightarrow \omega+\rmd u$.
The combination $g^{(z)} = \wt{E}\otimes E$, in particular, is then gauge invariant and globally-defined. We will call the quantity $g^{(z)} $ a metric, but we will \emph{not} be using this quantity to define areas or distances,\footnote{Indeed, it is well-known (see, e.g., \cite{Nelson88}) that using the notion of a metric to define ``area'' and ``distance'' on a~super Riemann surface is problematic. I am also grateful to Branislav Jurco for some correspondence on this.} which is what metrics are usually good for. Instead, what is important for our purposes is that~$g^{(z)}$ is globally defined, so that it can be used to specify a gauge slice to translate frames (and hence also NS punctures) across super Riemann surfaces in a well-defined and smooth manner.

For concreteness, we can actually proceed by direct analogy to the conformally-flat expression for the metric on an ordinary 2-sphere $\rmd \tilde{w}\rmd w/(1+\tilde{w}w)^2$ (where we chose the radius of the 2-sphere, $r=1/2$). Taking the aforementioned comments into account, consider the specific local expression for the ``metric''
\begin{gather}\label{eq:e^varphihat}
g^{(w)} = \frac{\rmd \tilde{w}(\rmd w-\rmd\psi\psi)}{(1+\tilde{w}w)^2}
\qquad{\rm with}\quad
{\rm e}^{\hat{\varphi}(\tilde{w};w|\psi)} = \frac{1}{(1+\wt{w}w)^2}.
\end{gather}
Note primarily that $g^{(w)}$ is not only gauge invariant (see Section~\ref{sec:SC} for some elaboration on this), it is also globally-defined in the sense that it is invariant under the superconformal transformation \eqref{eq:ztheta-S2-trans}, $g^{(u)}=g^{(w)}$.
One can check this by noting that under a general superconformal transformation $\wt{u};\!u|\theta_u\rightarrow \wt{w};\!w|\psi$, we have $\rmd \wt{w}=\rmd\wt{u}(\partial_{\tilde{u}}\wt{w})$ and $\varpi_w = \varpi_u(D_{\theta_u}\psi)^2$.

One may wonder to what extent such an expression for $g^{(w)}$ in \eqref{eq:e^varphihat} always exists on a~super Riemann surface with the topology of a 2-sphere. In fact, one can always arrive at this expression starting from any globally-defined (but otherwise arbitrary) metric $g^{(w')}$ of the correct topology, which may in turn involve any combination of odd and even variables (subject to the fact that~${\rm e}^{\varphi'(\tilde{w}';w'|\psi')}$ is even) by a unique (up to a phase) superconformal transformation $\wt{w}';\!w'|\psi'\rightarrow \wt{w};\!w|\psi$. A sketch of a proof that one can always map to a metric of the form \eqref{eq:e^varphihat} starting from an arbitrary metric is as follows: one can show this explicitly by building a Taylor series expansion for $\wt{w}(\wt{w}');\!w(w'|\psi')|\psi(w'|\psi')$ in terms of $\varphi'(\tilde{w}';w'|\psi')$ and its derivatives (evaluated at some base point of our choice). This Taylor series is guaranteed to be convergent since it is superconformal. The coefficients of this series expansion will, of course, not be superconformal in general (since they are constructed out of $\varphi'$ and its derivatives). (This calculation is similar to that elaborated on below, see \eqref{eq:TaylorTheta} and the associated discussion.)

Let us pause momentarily to make a brief remark on notation before we embark on the construction of the smooth gauge slice of interest. We denote the coordinate~$\tilde{w};\!w|\psi$ concisely by the superscript $^{\!\!(w)}$. A superscript $^{\!\!(z)}\!$ will similarly refer to the superconformal coordinate $\tilde{z};\!z|\theta$ and we reserve this notation for a special frame called a ``superconformal normal coordinate'' frame (that we will define momentarily). This will have the property that the origin ${\tilde{z};\!z|\theta=0;\!0|0}$ is identified with a supermodulus $\wt{v};\!v|\chi$ in the $^{\!\!(w)}$ coordinate system, which will in turn allow us to insert NS punctures at $\tilde{z};\!z|\theta=0;\!0|0$ and then translate them to integrated picture by mapping to the $^{\!\!(w)}$ coordinate system. We can then move the NS puncture across the super Riemann surface by allowing $\wt{v};\!v|\chi$ to vary, or, more precisely, by associating this quantity to a supermodulus. To simplify the notation, in the superconformal chart with coordinates~$\tilde{z};\!z|\theta$ we will sometimes write $\varphi^{(z)}$ as $\varphi$, and in the superconformally-related chart $\tilde{w};\!w|\psi$ we will occasionally write $\varphi^{(w)}$ as $\hat{\varphi}$, as in \eqref{eq:e^varphihat}. We also write $\hat{\varpi} = \rmd w-\rmd\psi\psi$ and $\varpi = \rmd z-\rmd\theta\theta$, and also $D_\theta=\partial_\theta+\theta\partial_z$.

\section{Superconformal normal coordinates}\label{sec:SCNC}
We now want to insert a NS puncture at some point on the super Riemann surface that in the~$^{(w)}$ coordinate with metric defined in \eqref{eq:e^varphihat}, will correspond to the coordinate ${\wt{w};\!w|\psi = \wt{v};\!v|\chi}$. So the~$2|1$ parameters $\wt{v};\!v|\chi$ will be identified with the even$|$odd moduli associated to this puncture. For example, we might like to insert an NS vertex operator at this point. We would like to use the standard operator/state correspondence of superconformal field theory on super Riemann surfaces, in which case it is natural to initially take this vertex operator to be in the $-1$ picture~\cite{FriedanMartinecShenker86,Witten12b}, and defined using radial quantisation on the flat super plane using a chart $\wt{z};\!z|\theta$, inserted at a~point with coordinate value $0;\!0|0$. So we start off with an NS vertex operator on a flat superplane in the $-1$ picture inserted at a point $\wt{z};\!z|\theta=0;\!0|0$. To transition to a global picture and associate this NS puncture location with a supermodulus we need to translate it to integrated picture. But we wish do so in a globally well-defined manner, and such that local super curvature is stored \emph{locally}. So we are generalising Polchinski's bosonic string construction~\cite{Polchinski88} to the superstring.\footnote{This is to be contrasted with the usual situation encountered in string field theory \cite{deLacroixErbinKashyapSenVerma17,Zwiebach93} (originally pioneered by Nelson \cite{Nelson89}), where this information that the sphere is curved is instead stored in transition functions on patch overlaps, see also \cite{LaNelson90}.}
After translating this vertex operator to integrated picture, we can then integrate over $\wt{v};\!v|\chi$ in the corresponding path integral.

The claim I would like to put forward now is the following. There exists a superconformal change of variables $\wt{w};\!w|\psi\rightarrow \wt{z};\!z|\theta$ that preserves the metric up to a superconformal factor,
\begin{gather}\label{eq:1/1wwbevarphi}
\frac{1}{(1+\tilde{w}w)^2} \rmd \tilde{w}(\rmd w-\rmd\psi\psi)=
{\rm e}^{\varphi(\tilde{z};z|\theta)}\rmd \tilde{z}(\rmd z-\rmd\theta\theta),
\end{gather}
such that at the location of the puncture, where $\wt{z};\!z|\theta=0;\!0|0$ (which maps to $\wt{w};\!w|\psi = \wt{v};\!v|\chi$), the new $g^{(z)}$ ``metric'' is ``as flat as possible'',
\begin{gather}\label{eq:SCNCgaugeslice}
 \varphi(0;\!0|0)=0 \quad{\rm and}\qquad D_\theta^n\varphi(0;\!0|0) =\partial_{\tilde{z}}^n \varphi(0;\!0|0)=0,\qquad n\geq1.
\end{gather}
The presence of super curvature (which in the $^{(z)}$ frame reads $\EuScript{R}_{\tilde{z}\theta}=-\partial_{\tilde{z}}D_\theta \varphi$), see Section~\ref{sec:SC}, means that mixed derivatives cannot be set to zero by a superconformal transformation, but there is no obstruction to setting purely holomorphic or purely anti-holomorphic derivatives of~$\varphi$ equal to zero at a point as done in~\eqref{eq:SCNCgaugeslice}.

I would like to suggest that \eqref{eq:SCNCgaugeslice} is the appropriate generalisation to the heterotic string of the gauge slice constructed in the context of bosonic string theory by Polchinski in \cite{Polchinski88}. We will next prove that this gauge slice \eqref{eq:SCNCgaugeslice} exists and in the process we will also derive the precise superconformal transformation that implements it. It will turn out to be a specific $\mathrm{OSp}(2,1)$ transformation\footnote{The group $\mathrm{OSp}(2,1)$ is referred to as $\mathrm{SPL}(2,\mathbb{C})$ in \cite{CraneRabin88}.} with parameters that depend smoothly on $\wt{v};\!v|\chi$.

We will build this superconformal transformation using a Taylor series expansion for $\wt{z}(\wt{w})$, $\theta(w|\psi)$ and $z(w|\psi)$. A Taylor series for $\theta(w|\psi)$ about $\wt{w};\!w|\psi = \wt{v};\!v|\chi$ (taking into account that $\theta(v|\chi)=0$) takes the general form (see \cite[Section~2.7]{Friedan86})
\begin{gather}\label{eq:TaylorTheta}
\theta(w|\psi)
= \hat{\psi}D_\psi\theta(v|\chi) +\sum_{n=1}^\infty\frac{1}{n!}\hat{w}^n
\big(\partial_w^n\theta(v|\chi)+\hat{\psi} \partial_w^nD_\psi\theta(v|\chi)\big),
\end{gather}
with $\hat{\psi}=\psi-\chi$ and $\hat{w}=w-v-\psi\chi$. What distinguishes one superconformal transformation from another are the coefficients $\partial_w^nD_\psi\theta$ and $\partial_w^n\theta$. In particular, we need to derive explicit expressions for $\partial_w^nD_\psi\theta$ and $\partial_w^n\theta$ for $n=0,1,\dots$ subject to the gauge slice conditions \eqref{eq:SCNCgaugeslice}. Using the superconformal chain rule (according to which $D_\psi=(D_\psi\theta)D_\theta$, etc.), it is immediate to see that if \eqref{eq:SCNCgaugeslice} is satisfied at $\wt{z};\!z|\theta=0;\!0|0$ then so will $D_\psi^n\varphi=\partial_{\tilde{w}}^n\varphi=0$ be satisfied at~${\wt{z};\!z|\theta=0;\!0|0}$, and vice versa. Since furthermore we require the ``metric'' to be globally-defined, under superconformal transformations, we wish to set $g^{(z)}=g^{(w)}$, as indicated in \eqref{eq:1/1wwbevarphi}. This implies the superconformal factor transforms as
\begin{gather}\label{eq:varphiwvarphiz}
\varphi^{(w)} = \varphi^{(z)}-\ln(D_\theta\psi)^2-\ln\partial_{\tilde{z}}\tilde{w},
\end{gather}
with $\hat{\varphi}\equiv \varphi^{(w)}$ in the case of interest given by the sphere metric \eqref{eq:e^varphihat}.
We then hit \eqref{eq:varphiwvarphiz} with an appropriate number of derivatives, $\partial_w$, $D_\psi$, and evaluate the resulting relations at $\wt{z};\!z|\theta=0;\!0|0$ (equivalently, $\wt{w};\!w|\psi = \wt{v};\!v|\chi$). After some elementary manipulations, we find
\begin{gather}
\partial_w^n\theta = D_\psi \bigl(B_{n-1}\bigl(\tfrac{1}{2}\partial_w^s\h{\varphi}\bigr)D_\psi\theta\bigr),\qquad n>1,\qquad
\partial_w\theta=\tfrac{1}{2}D_\psi\h{\varphi} D_\psi\theta,\nonumber\\[1mm]
\partial_w^nD_\psi \theta = B_{n}\bigl(\tfrac{1}{2}\partial_w^s\h{\varphi}\bigr)D_\psi\theta, \qquad n>0,\label{eq:TaylorIdentities}
\end{gather}
which are evaluated at the location of the puncture. The quantities $B_n(a_s)\equiv B_n(a_1,\dots,a_n)$ are complete Bell polynomials. We will keep the argument $a_s=\tfrac{1}{2}\partial_w^s\h{\varphi}(\wt{v};\!v|\chi)$ implicit for conciseness. The relations \eqref{eq:TaylorIdentities} follow from \eqref{eq:SCNCgaugeslice}, \eqref{eq:varphiwvarphiz}, and standard properties of complete Bell polynomials. We then substitute the relations \eqref{eq:TaylorIdentities} back into the Taylor expansion formula~\eqref{eq:TaylorTheta} to arrive at
\begin{gather}\label{eq:thetawpsia0}
\theta(w|\psi)=D_\psi\theta(v|\chi)
\left\{\h{\psi}+\sum_{n=1}^\infty\frac{1}{n!}\h{w}^n\left(D_\psi B_{n-1}+\frac{1}{2}D_\psi \h{\varphi}B_{n-1}+\h{\psi}B_{n}\right)(\wt{v};\!v|\chi)\right\}.
\end{gather}
We next evaluate the various complete Bell polynomials making use of the sphere metric \eqref{eq:e^varphihat},
\begin{gather*}
D_\psi B_{n-1} = (n-1)(n-1)\left(\frac{-\wt{v}}{1+\wt{v}v}\right)^n\chi,\qquad
\frac{1}{2}D_\psi \h{\varphi}B_{n-1} = (n-1)\left(\frac{-\wt{v}}{1+\wt{v}v}\right)^n\chi,\\
\h{\psi}B_{n}= n\left(\frac{-\wt{v}}{1+\wt{v}v}\right)^n(\psi-\chi),
\end{gather*}
and substitute these into \eqref{eq:thetawpsia0}. Carrying out the sum over $n$ and rearranging, we arrive at
\begin{gather}\label{eq:thetawpsia}
\theta(w|\psi)=D_\psi\theta(v|\chi)\frac{(1+\wt{v}v)(\psi-\chi)-\wt{v}(w-v)\chi}{\wt{v}w+1}.
\end{gather}
We have not yet determined $D_\psi\theta(v|\chi)$. Putting this aside momentarily, we next extract the corresponding expression for $z(w|\psi)$.

A simple way to construct a Taylor series for $z(w|\psi)$ is to substitute \eqref{eq:thetawpsia} into the superconformal condition $D_\psi z = \theta D_\psi \theta$ and integrate it using the boundary condition $z(v|\chi) = 0$ (which is inherited from the fact that the puncture is inserted at $\wt{z};\!z|\theta = 0$). This procedure leads to the explicit expression
\begin{gather}\label{eq:zwpsia}
z(w|\psi)=[D_\psi\theta(v|\chi)]^2\frac{(1+\wt{v}v)(w-v-\psi\chi)}{\wt{v}w+1}.
\end{gather}

Notice that despite the fact that the transition functions \eqref{eq:thetawpsia} and \eqref{eq:zwpsia} are superconformal in $w|\psi$, they are nevertheless only \emph{smooth} in the supermoduli $\wt{v};\!v|\chi$.

Carrying out the same procedure for the anti-chiral half, the relation analogous to \eqref{eq:TaylorIdentities} is (see \cite[Section~2.4.2]{SklirosLuest21})
$
\partial_{\tilde{w}}^n\tilde{z} = B_{n-1}\big(\partial_{\tilde{w}}^s\hat{\varphi}\big),
$
which is evaluated at $\wt{v};\!v|\chi$,
and therefore the Taylor series expansion for $\tilde{z}(\tilde{w})$ around $\wt{w}=\wt{v}$ reads
\begin{gather}\label{eq:ztilde(p)-existence}
\tilde{z}(\tilde{w})=\partial_{\tilde{w}}\tilde{z}(\wt{v})\left\{\sum_{n=1}^\infty\frac{1}{n!}(\tilde{w}-\tilde{v})^n B_{n-1}\big(\partial_{\tilde{w}}^s\h{\varphi}\big)\right\}.
\end{gather}
Evaluating the complete Bell polynomial taking into account \eqref{eq:e^varphihat} leads to
\[
B_{n-1}\big(\partial_{\tilde{w}}^s\h{\varphi}\big) = n!\big(\frac{-v}{1+\wt{v}v}\big)^{n-1},\]
which in turn implies that (carrying out the sum over $n$) \eqref{eq:ztilde(p)-existence} reduces to
\begin{gather}\label{eq:ztildewt}
\wt{z}(\wt{w}) = \partial_{\tilde{w}}\tilde{z}(\wt{v})\frac{(1+\wt{v}v)(\wt{w}-\wt{v})}{v\wt{w}+1}.
\end{gather}

Returning now to the quantity $D_\psi\theta(v|\chi)$ in \eqref{eq:thetawpsia} or \eqref{eq:zwpsia}, and also now $\partial_{\tilde{w}}\wt{z}(\wt{v})$ in \eqref{eq:ztildewt}, these are in fact not independently determined by the gauge slice. Instead, it is only the combination
$\partial_{\tilde{w}}\wt{z}(\wt{v}) (D_\psi\theta(v|\chi))^2=1/(1+\wt{v}v)^2$
that is determined. As in the bosonic string \cite{Polchinski88}, there is an obstruction (the Euler number) to setting the phase of $\theta$ or $z$ to zero globally, but this will be sufficient. (For a detailed derivation of this point see \cite[Section~2.5]{SklirosLuest21} and in particular the discussion on \cite[p.~46]{SklirosLuest21}, and also Section~2.4.2 therein.) This is the topological or global origin of the $L_0-\wt{L}_0=0$ constraint that must be satisfied by local vertex operators in superstring perturbation theory \cite{Nelson89,Polchinski88} or string fields in string field theory \cite{deLacroixErbinKashyapSenVerma17,Zwiebach93}. Notice that the superconformal condition implies $(D_\psi\theta)^2=\partial_wz-\partial_w\theta \theta$, and since $\theta(v|\chi)=0$ the ambiguity is identical to that in the bosonic string \cite{Polchinski88,SklirosLuest21}, namely
\begin{gather}\label{eq:dwztdwz}
\partial_{\tilde{w}}\wt{z}(\wt{v}) \partial_wz(v|\chi)=\frac{1}{(1+\wt{v}v)^2}.
\end{gather}
This is as expected, because (apart from the notion of a spin structure) there is no new non-trivial topological information on the worldsheet that arises due to the presence of odd variables (see~\cite[Section~2.1.2]{Witten12a}). From \eqref{eq:dwztdwz} we see that there can be no $\chi$ dependence in either~$\partial_wz(v|\chi)$ or~$\partial_{\tilde{w}}\wt{z}(\wt{v})$, because there is no other odd variable present in \emph{these} transition functions. (There can still be other odd moduli associated to other punctures or handle operators inserted elsewhere on the super Riemann surface.) Requiring that these become complex conjugates when we set odd variables to zero \cite{Witten12a} then determines each of these up to a $v$,$\wt{v}$-dependent phase
\begin{gather*}
\partial_{\tilde{w}}\wt{z}(\wt{v}) = \frac{{\rm e}^{-{\rm i}\alpha(\tilde{v},v)}}{1+\wt{v}v},\qquad \partial_wz(v|\chi) = \frac{{\rm e}^{{\rm i}\alpha(\tilde{v},v)}}{1+\wt{v}v},\qquad{\rm and}\qquad D_\psi\theta(v|\chi) = \frac{\pm {\rm e}^{\frac{{\rm i}}{2}\alpha(\tilde{v},v)}}{\sqrt{1+\wt{v}v}}.
\end{gather*}
The phase $\alpha(\tilde{v},v)$ is real when we set $\wt{v}=v^*$ (where $v^*$ is the complex conjugate of $v$). The `$\pm$' sign in $D_\psi\theta(v|\chi)$, denoted by $\eta$ in what follows, is meaningful and is associated to a choice of spin structure.
Although $\alpha$ does depend on $\wt{v}$, $v$, it must always cancel out of observable quantities. It will therefore be convenient to absorb $\alpha$ into a redefinition $\wt{z};\!z|\theta\rightarrow {\rm e}^{{\rm i}\alpha}\wt{z};\!{\rm e}^{-{\rm i}\alpha}z|{\rm e}^{-{\rm i}\alpha/2}\theta$ and instead check explicitly that the physically-meaningful quantities do not depend on such a phase.

Summarising, the superconformal transformation $\wt{w};\!w|\psi\rightarrow \wt{z};\!z|\theta$ that maps the globally-defined sphere coordinate $\wt{w};\!w|\psi$ to the flat superplane\footnote{By ``flat superplane coordinate'' we mean the superconformal frame that is associated to the metric $g^{(z)} = {\rm e}^\varphi\rmd \tilde{z}(\rmd z-\rmd\theta\theta)$ which satisfies $\varphi=0$ (and in particular \eqref{eq:SCNCgaugeslice}) at the puncture $\wt{z};\!z|\theta=0;\!0|0$.} coordinate, $\wt{z};\!z|\theta$, that in turn translates a NS puncture inserted at $\wt{w};\!w|\psi=\wt{v};\!v|\chi$ to the point $\wt{z};\!z|\theta=0;\!0|0$ is given by
\begin{gather}
\wt{z}(\wt{w}) = \frac{\wt{w}-\wt{v}}{v\wt{w}+1},\qquad
z(w|\psi)=\frac{w-v-\psi\chi}{\wt{v}w+1},\qquad
\theta(w|\psi)
=\eta \frac{\sqrt{1+\wt{v}v}}{(\wt{v}w+1)}\psi-\eta \frac{\chi}{\sqrt{1+\wt{v}v}},\!\!\!
\label{eq:superconftrans}
\end{gather}
where $\eta=\pm1$. We also need the inverses
\begin{gather}
\wt{w}(\wt{z})= \frac{\wt{z}+\wt{v}}{-v\wt{z}+1},\qquad
w(z|\theta) = \frac{z+v+\eta\theta\chi/\sqrt{1+\wt{v}v}}{-\wt{v}z+1-\eta\theta\chi \wt{v}/\sqrt{1+\wt{v}v}},\nonumber\\
\psi(z|\theta)= \frac{\sqrt{1+\wt{v}v} \eta\theta+\chi}{-\wt{v}z+1},\label{eq:wtwpsiOSp}
\end{gather}
which follow from \eqref{eq:superconftrans}.
By construction, notice that $\wt{w}=w^*$ when we set the odd variables equal to zero, but we have not required any stronger version of complex conjugation.

The quantities \eqref{eq:superconftrans} and \eqref{eq:wtwpsiOSp} define the notion of a specific super frame $E_A^{\phantom{A}M}$ and its inverse~$E_M^{\phantom{M}A}$, respectively, where $A=\wt{z},z,\theta$ denotes the frame indices and $M=\wt{w},w,\psi$ could be thought of as Einstein (or base coordinate) indices. At $\wt{w};\!w|\psi=\wt{v};\!v|\chi$ (equivalently $\wt{z};\!z|\theta=0;\!0|0$), these read explicitly
\begin{gather*}
E_M^{\phantom{A}A} =
\left(
 \begin{matrix}
 E_{\tilde{w}}^{\phantom{w}\tilde{z}} & E_{\tilde{w}}^{\phantom{w}z}& E_{\tilde{w}}^{\phantom{w}\theta} \\
 E_{w}^{\phantom{w}\tilde{z}} & E_{w}^{\phantom{w}z} & E_{w}^{\phantom{w}\theta} \\
 E_{\psi}^{\phantom{\psi}\tilde{z}}& E_{\psi}^{\phantom{\psi}z}& E_{\psi}^{\phantom{\psi}\theta}\\
 \end{matrix}
\right)
=
\left(
 \begin{matrix}
 \frac{1}{1+\tilde{v}v} & 0& 0 \\
 0 & \frac{1}{1+\tilde{v}v}& -\frac{\eta\chi\tilde{v}}{(1+\tilde{v}v)^{3/2}}\\
 0& -\frac{\chi}{1+\tilde{v}v}& \frac{\eta}{\sqrt{1+\tilde{v}v}}\\
 \end{matrix}
\right),\\
E_A^{\phantom{A}M} =
\left(
 \begin{matrix}
 E_{\tilde{z}}^{\phantom{a}\tilde{w}} & E_{\tilde{z}}^{\phantom{a}w}& E_{\tilde{z}}^{\phantom{w}\psi} \\
 E_{z}^{\phantom{a}\tilde{w}} & E_{z}^{\phantom{a}w} & E_{z}^{\phantom{a}\psi} \\
 E_{\theta}^{\phantom{a}\tilde{w}}& E_{\theta}^{\phantom{a}w}& E_{\theta}^{\phantom{a}\psi}\\
 \end{matrix}
\right)
=
\left(
 \begin{matrix}
 1+\tilde{v}v & 0& 0 \\
 0 &1+\tilde{v}v& \chi\tilde{v}\\
 0& \eta\chi\sqrt{1+\tilde{v}v}& \eta\sqrt{1+\tilde{v}v}\\
 \end{matrix}
\right).
\end{gather*}
The individual entries are defined as expected, e.g., $E_{w}^{\phantom{w}z}=\frac{\partial z}{\partial w}|_{w=v}$, $E_{\psi}^{\phantom{w}z}=\frac{\partial z}{\partial \psi}|_{w=v}$, etc., whereas the corresponding Berezinian \cite{Witten12a} for the change of coordinates (evaluated at the puncture) is given by
\begin{gather}\label{eq:DDBer}
\mathcal{D}(\wt{z},z|\theta) = \mathcal{D}(\wt{v},v|\chi)\Ber E_M^{\phantom{a}A}\qquad{\rm with}\quad \Ber E_M^{\phantom{a}A} = \frac{\eta}{(1+\wt{v}v)^{3/2}}.
\end{gather}

\section{Super curvature}\label{sec:SC}
It proves useful, especially in the case of more general super Riemann surfaces to introduce the notion of \emph{super curvature}. Rather than provide the details of this general discussion here however, we will instead only present, very briefly, the ingredients we will be needing in the current paper.

We can define super curvature of a heterotic super Riemann surface $\Sigma$ as follows.
Following~\cite{Witten12b}, we first decompose the contangent bundle $T^*\Sigma=T^*_L\Sigma\oplus T^*_R\Sigma$ (where $T^*_L\Sigma$ is of rank~$1|0$ and $T^*_R\Sigma$ is of rank $1|1$) by declaring that $T^*_L\Sigma$ is generated by a quantity $\wt{E}$, whereas $T^*_R\Sigma$ is generated by $E$ and $F$ (see \cite[Section~3.6 and in particular Section~3.6.3]{Witten12b}). The quantity $E$ generates a subbundle $\mathcal{D}^{-2}\subset T^*_R\Sigma$.
We then introduce a connection $\omega$ on the line bundle $\mathcal{D}^{-2}$ and postulate a gauge invariance,
\begin{gather*}
\wt{E}\rightarrow {\rm e}^{-u}\wt{E},\qquad E\rightarrow {\rm e}^uE,\qquad F\rightarrow {\rm e}^{u/2}F,\qquad \omega\rightarrow \omega+\rmd u.
\end{gather*}
The corresponding gauge-covariant exterior derivatives of $\wt{E}$, $E$ and $F$ are then
\[
\D\wt{E}= (\rmd +\omega)\wt{E},\qquad
\D E = (\rmd -\omega)E,\qquad
\D F=\left(\rmd -\frac{1}{2}\omega\right)F,
\]
where $\rmd = \rmd \wt{z} \partial_{\tilde{z}}+\varpi \partial_z+\rmd \theta D_\theta$ is the ordinary exterior derivative, and a convenient component expansion for $\omega$ in the chart $(\U_{\bm{z}},\wt{z};\!z|\theta)$ is then $\omega = \rmd \wt{z} \omega_{\tilde{z}}+\varpi \omega_z+\rmd\theta  \omega_\theta$. The super analogues of \emph{metric compatibility} and \emph{vanishing torsion} are encoded in
$
\D \wt{E} =0$,
$\D E+F\wedge F= 0$.
In practice, it is convenient to fix the above gauge invariance. Omitting details, in a local chart~$^{(z)}$ this analysis leads to the following explicit expressions:
\begin{gather*}
\wt{E}^{(z)}={\rm e}^{\varphi}\rmd\wt{z},\qquad E^{(z)}= \varpi,\qquad F^{(z)}=\rmd\theta+\varpi\frac{1}{2}D_\theta\varphi,
\end{gather*}
where $\varphi(\wt{z};\!z|\theta)$ is a smooth function of the arguments (but may implicitly also depend smoothly on supermoduli). Notice that the ``metric'' we defined in Section~\ref{sec:SCNC} (and as very briefly mentioned in Section~\ref{sec:SRS}), namely $g^{(z)}= {\rm e}^\varphi \rmd\wt{z}\otimes \varpi$, is none other than the gauge-invariant combination $\wt{E}\otimes E$ after gauge fixing.\footnote{We could have included a term $\lambda \wt{E}\otimes F^2$ in $g^{(z)}$ with $\lambda$ an odd smooth function of $\wt{z};\!z|\theta$; this would also be gauge-invariant, but it does not seem necessary since our expression for $g^{(z)}$ is already globally-defined and gauge-invariant.} The gauge-fixed expression for the connection, $\omega$, in turn reads
\begin{gather*}
\omega = \rmd \wt{z}\omega_{\tilde{z}}+\varpi \omega_z +\rmd\theta\omega_\theta\qquad\text{with}\quad
\omega_{\tilde{z}} = 0,\qquad
\omega_z  = -\partial_z\varphi,\qquad
\omega_{\theta}  = -D_\theta\varphi.
\end{gather*}

We can use these quantities to arrive at a useful notion of \emph{super curvature}, ${\R}$, defined by
\begin{gather*}
\D^2 = n\R\qquad{\rm with}\quad \R = \varpi\rmd\wt{z} \R_{z\tilde{z}}+\rmd\wt{z}\rmd\theta {\R}_{\tilde{z}\theta},
\end{gather*}
and $n$ is the $\mathrm{U}(1)$ weight of the superconformal tensor on which the operator $\D^2$ acts. We have taken into account that $\D$ preserves $\mathrm{U}(1)$ weight, and have defined the quantities ${\R}_{z\tilde{z}}\equiv-D_\theta {\R}_{\tilde{z}\theta}$ and
$
{\R}_{\tilde{z}\theta} \equiv -\partial_{\tilde{z}}D_\theta \varphi$.
We will usually refer to the component ${\R}_{\tilde{z}\theta}$ (out of which the entire expression for $\R$ can be reconstructed) as the \emph{super curvature}. Under superconformal changes of coordinates (i.e., analytic maps $\wt{z};\!z|\theta\rightarrow \wt{w}(\wt{z});\!w(z|\theta)|\psi(z|\theta)$ subject to $D_\theta w = \psi D_\theta\psi$) it transforms as a section of $(T_L^*\Sigma)\otimes \mathcal{D}^{-1}$, in particular,
$
{\R}_{\tilde{z}\theta}={\R}_{\tilde{w}\psi}(\partial_{\tilde{z}}\tilde{w})(D_\theta\psi)$,
the corresponding~$\mathrm{U}(1)$ weight being therefore $n=-\frac{1}{2}$. Super curvature thus transforms as an odd smooth section of the Berezinian. The quantity $\R$ is globally-defined, in the sense that under superconformal transformations it transforms as
$
\R^{(z)} = \R^{(w)}.
$
We will often find it convenient to work in terms of the super curvature, ${\R}_{\tilde{z}\theta}$, and its derivatives, since it is these quantities that appear in the path integral measure and vertex operators in heterotic string theory.

If we now focus on the specific super Riemann surface of interest, namely that with the topology of a 2-sphere, in the $^{(w)}$ superconformal frame (defined in Section~\ref{sec:SRS}) the super curvature is given by the local expression $\EuScript{R}_{\tilde{w}\psi}= -\partial_{\tilde{w}}D_{\psi}\hat{\varphi}$, so that according to \eqref{eq:e^varphihat}
\begin{gather}\label{eq:Rwtpsi}
\EuScript{R}_{\tilde{w}\psi}
=\frac{2\psi}{(1+\tilde{w}w)^2}.
\end{gather}
Since super curvature is an odd smooth section of the Berezinian\footnote{We are being somewhat heuristic here, a more general discussion will be presented elsewhere.}
 $\EuScript{R}_{\tilde{w}\psi} = (\partial_{\tilde{z}}\tilde{w} D_\theta\psi)^{-1}\EuScript{R}_{\tilde{z}\theta}$ (the integration measure $\mathcal{D}(\tilde{w};\!w|\psi) =\mathcal{D}(\tilde{z};\!z|\theta)\partial_{\tilde{z}}\tilde{w} D_\theta\psi$ transforms in the opposite manner), the integral
$
\chi = \frac{1}{2\pi}\int\mathcal{D}(\tilde{w};\!w|\psi)\EuScript{R}_{\tilde{w}\psi}$
is well-defined and in fact equals the \emph{Euler characteristic} of the super Riemann surface. We leave it as an exercise for the reader to check that
the Euler characteristic is given by its classical value $\chi=2$ as expected \cite{Witten12b}. An important point is that (on the super sphere) we only need a~single coordinate chart to compute this quantity, because we incorporate super curvature locally and furthermore it dies off sufficiently rapidly at infinity.

The super curvature $\EuScript{R}_{\tilde{w}\psi}$ given in \eqref{eq:Rwtpsi} is in the $^{(w)}$ frame. We will also need the corresponding expression in the $^{(z)}$ frame. Since the two frames are related by $\EuScript{R}_{\tilde{z}\theta} = \partial_{\tilde{z}}\wt{w}D_\theta\psi\EuScript{R}_{\tilde{w}\psi}$, according to \eqref{eq:wtwpsiOSp} and \eqref{eq:Rwtpsi},
\begin{gather*}
\EuScript{R}_{\tilde{z}\theta}(z;\!z|\theta) = \frac{2\theta}{(1+\tilde{z}z)^2}+\frac{2\eta\chi}{\sqrt{1+\tilde{v}v}(1+\tilde{z}z)^2},
\end{gather*}
and therefore at $\wt{z};\!z|\theta=0;\!0|0$ (equivalently $\wt{w};\!w|\psi= \wt{v};\!v|\chi$)
\begin{gather}\label{eq:supercurv}
\EuScript{R}_{\tilde{z}\theta} = \frac{2\eta\chi}{\sqrt{1+\wt{v}v}}\qquad{\rm and}\qquad
\frac{1}{2}D_\theta\EuScript{R}_{\tilde{z}\theta}=1,
\end{gather}
with the following (purely chiral or purely anti-chiral) higher derivatives vanishing: $\partial_z^nD_\theta\EuScript{R}_{\tilde{z}\theta}=\partial_z^n\EuScript{R}_{\tilde{z}\theta}=\partial_{\tilde{z}}^n\EuScript{R}_{\tilde{z}\theta}=0$ for all $n=1,2,\dots$. Incidentally, there is also a notion of `torsion' on super Riemann surfaces \cite{DHokerPhong}. Since we will not be making explicit use of this below, for completeness we simply mention that the torsion constraints are automatically satisfied in the gauge slice of interest, and in particular we find
$\T_{\theta\theta}^{\phantom{aa}z}=2$ and $
\T_{z\tilde{z}}^{\phantom{aa}\theta}=-\frac{1}{2}\R_{\tilde{z}\theta}$
with all remaining components equal to zero.

\section{Path integral measure}\label{sec:PIM}
To implement the gauge slice developed in Section~\ref{sec:SCNC} into the corresponding path integral, we need to determine the path integral measure. For this, we will need to know the change in~$\wt{z}(\wt{w})$, $z(w|\psi)$, $\theta(w|\psi)$ with respect to small variations in the supermoduli $\wt{v};\!v|\chi$ keeping the coordinate $\wt{w};\!w|\psi$ fixed. In fact, by keeping $\wt{w};\!w|\psi$ fixed we are also keeping the underlying metric fixed.\footnote{This hint provides the starting point towards understanding how to carry out the corresponding computation for arbitrary super Riemann surfaces with arbitrary super curvature (subject to the Euler number constraint). It will be discussed elsewhere.} This is because (as seen in \eqref{eq:e^varphihat}) in the $^{(w)}$ coordinate system the metric depends solely on the coordinates $\wt{w};\!w|\psi$. (This is to be contrasted with $\varphi^{(z)}$ which will also depend on supermoduli.)
The explicit expression for the path integral insertion that will implement our gauge slice is
\begin{gather}\label{eq:NSpuncmeasure}
\int \mathcal{D}(\wt{v},\!v|\chi)\delta\big(\h{\EuScript{B}}_{\tilde{v}}\big)\delta\big(\h{\EuScript{B}}_{v}\big)\delta\big(\h{\EuScript{B}}_{\chi}\big),
\end{gather}
where we adopt the shorthand $\mathcal{D}(\wt{v},\!v|\chi)=-{\rm i}[\rmd \wt{v},\!\rmd v|\rmd\chi]$. (Note that the action also depends on supermoduli, and so this must be included when we actually integrate over them.) This quantity \eqref{eq:NSpuncmeasure} acts on a single fixed-picture NS vertex operator in the $-1$ picture (that may or may not be offshell) defined in the $^{(z)}$ frame and inserted at $\wt{z};\!z|\theta=0;\!0|0$. Let $t$ stand for any of the quantities $\wt{v}$, $v$ or $\chi$. Since there is only a single patch overlap $\U_z\cap \U_w$ (which in turn corresponds to an annulus or a punctured disc with the origin $\wt{z};\!z|\theta=0;\!0|0$ absent), the superghost insertions appearing in \eqref{eq:NSpuncmeasure} are then determined from
\begin{gather}\label{eq:Bthat3}
\h{\EuScript{B}}_t=\frac{1}{2\pi {\rm i}}\int_{C_{zw}}\left(-[\dee z|\dee \theta]\bigg[\frac{\partial z}{\partial t}-\frac{\partial \theta}{\partial t}\theta\bigg]_{w|\psi}B_{z\theta}+(-)^{|t|}\dee \wt{z}\left[\frac{\partial \tilde{z}}{\partial t}\right]_{\tilde{w}}\wt{b}_{\tilde{z}\tilde{z}}\right),
\end{gather}
which can be derived from the expression for the measure given by Witten in \cite{Witten12c} by a procedure precisely analogous to that in the bosonic string as derived in \cite[Section~9]{Polchinski_v1} or in \cite[Section~3]{SklirosLuest21}. The derivation linking the two viewpoints is in particular precisely analogous to the derivation linking the first and second equalities in \cite[equation~(3.245)]{SklirosLuest21}, but we will omit the details. See also \cite{WangYin22} for some further context and a more complete discussion.

The contour $C_{zw}$ in \eqref{eq:Bthat3} traverses the annular overlap $\U_z\cap \U_w$ enclosing the origin $\wt{z};\!z|\theta=0;\!0|0$ in a counterclockwise sense from the viewpoint of $\U_z$, so that
\[
\int_{C_{zw}}[\rmd z|\rmd\theta]\theta/z=-\int_{C_{zw}}\rmd\wt{z}/\tilde{z}=2\pi {\rm i}.
\]
 We define $|t|$ to be 0 or 1 for $t$ Grassmann-even or odd parity, respectively.\footnote{It might be useful to display the even and odd Grassmann parity quantities \smash{$\big|\wt{v}\big|=|v|=\big|\h{\EuScript{B}}_\chi\big|=0$} and \smash{$|\chi|=\big|\h{\EuScript{B}}_{\tilde{v}}\big|=\big|\h{\EuScript{B}}_{v}\big|=1$}, respectively.}

The notation for the derivatives appearing in \eqref{eq:Bthat3} indicates that we differentiate the frame coordinates $\wt{z}$, $z$, $\theta$ in \eqref{eq:superconftrans} with respect to the supermoduli $t=\wt{v}$, $v$ or $\chi$ keeping $\wt{w};\!w|\psi$ fixed. After taking these derivatives, we will make use of the inverse expressions given in \eqref{eq:wtwpsiOSp} to~eliminate the $\wt{w};\!w|\psi$ dependence in favour of $\wt{z}$, $z$, $\theta$ (since the contour integrals in \eqref{eq:Bthat3} are over~$\wt{z}$,~$z$,~$\theta$, and furthermore the superghosts are also defined using the $^{(z)}$ frame). Another~techni\-cal detail is that there is some information about the phase of $z+\delta z(z|\theta)$ with~$\delta z(z|\theta)$ generated by the aforementioned supermoduli variations. This $\wt{v}$,$v$-dependent phase \smash{${\rm e}^{2{\rm i} \operatorname{Im} \frac{\tilde{v}\delta v}{1+\tilde{v}v}}$} is not physically meaningful, so we can set it to zero provided we can show that physical observables do not depend on it, and there is a similar remark for the phase \smash{${\rm e}^{{\rm i} \operatorname{Im} \frac{\tilde{v}\delta v}{1+\tilde{v}v}}$} of $\theta+\delta \theta(z|\theta)$ (see Appendix~\ref{sec:MC}).

A short computation (see Appendix \ref{sec:MC}) implementing the above procedure then leads to the following results for the derivatives appearing in \eqref{eq:Bthat3}. In terms of the quantities
\begin{gather*}
\V_t(z|\theta)\equiv \left[\frac{\partial z}{\partial t}-\frac{\partial \theta}{\partial t}\theta\right]_{w|\psi}
\qquad{\rm and}\qquad
\wt{\V}_{t}(\wt{z})\equiv \left[\frac{\partial \wt{z}}{\partial t}\right]_{\tilde{w}},
\end{gather*}
one finds (for $t=\wt{v}$, $v$ or $\chi$) in particular
\begin{alignat}{3}
&\V_{\tilde{v}}(z|\theta) =-\frac{1}{1+\wt{v}v}\left(z^2-\frac{2\eta\chi}{\sqrt{1+\wt{v}v}}z\theta\right),\qquad&&   \wt{\V}_{\tilde{v}}(\wt{z}) =-\frac{1}{1+\wt{v}v},&\nonumber\\
&\V_v(z|\theta)= -\frac{1}{1+\wt{v}v}\left(
1+\frac{ 2\eta\chi }{\sqrt{1+\wt{v}v}}\wt{v}\theta\right),\qquad&&  \wt{\V}_v(\wt{z})= -\frac{\wt{z}^2}{1+\wt{v}v},&\nonumber\\
&\V_{\chi} (z|\theta)=\frac{\eta}{\sqrt{1+\wt{v}v}}
\left(2\theta+\frac{\eta\chi}{\sqrt{1+\wt{v}v}}\right),\qquad&& \wt{\V}_{\chi}(\wt{z}) =0.&\label{eq:VtVcomp}
\end{alignat}
Notice that the dependence of these quantities on $z|\theta$ and $\wt{z}$ is, respectively, \emph{superconformal}, but the dependence on the supermoduli $\wt{v};\!v|\chi$ is instead only \emph{smooth}.
Substituting these expressions~\eqref{eq:VtVcomp} into \eqref{eq:Bthat3} while taking into account the contour integral representations for the~$^{(z)}$ frame superghost modes
\begin{gather*}
\wt{b}_n^{(z)}=-\frac{1}{2\pi {\rm i}}\oint [\rmd \tilde{z}]\tilde{z}^{n+1}\wt{b}_{\tilde{z}\tilde{z}}(\tilde{z}),\qquad
b_{n}^{(z)}=\frac{1}{2\pi {\rm i}}\oint [\dee z|\dee \theta]z^{n+1}B_{z\theta}(z|\theta),\\
\beta_{n+1/2}^{(z)}=\frac{1}{2\pi {\rm i}}\oint [\dee z|\dee \theta]\theta z^{n+1}B_{z\theta}(z|\theta)
\end{gather*}
leads to the following explicit expressions for the measure
\begin{gather}
\h{\EuScript{B}}_{\tilde{v}}
=\frac{1}{1+\wt{v}v}\left(\wt{b}_{-1}^{(z)}+b_1^{(z)}+\frac{2\eta\chi}{\sqrt{1+\wt{v}v}}\beta_{1/2}^{(z)}\right),\nonumber\\
\h{\EuScript{B}}_{v}
=\frac{1}{1+\wt{v}v}\left(b_{-1}^{(z)}+\wt{b}_1^{(z)}-\frac{ 2\eta\chi \wt{v}}{\sqrt{1+\wt{v}v}}\beta_{-1/2}^{(z)}\right),\nonumber\\
\h{\EuScript{B}}_{\chi}
=\frac{\eta}{\sqrt{1+\wt{v}v}} \left(-2\beta_{-1/2}^{(z)}+\frac{\eta\chi}{\sqrt{1+\wt{v}v}}b_{-1}^{(z)}\right),\label{eq:BBB}
\end{gather}
and in particular the full insertion \eqref{eq:NSpuncmeasure} takes the form
\begin{gather}
\int \mathcal{D}(\wt{v},\!v|\chi){\rm e}^{-I_{\tilde{v}v\chi}}\delta\big(\h{\EuScript{B}}_{\tilde{v}}\big)\delta\big(\h{\EuScript{B}}_{v}\big)\delta\big(\h{\EuScript{B}}_{\chi}\big)\nonumber\\
\qquad=\eta\int \mathcal{D}(\wt{v},\!v|\chi)(1+\wt{v}v)^{-3/2}{\rm e}^{-I_{\tilde{v}v\chi}}\left(\wt{b}_{-1}^{(z)}+b_1^{(z)}+\frac{2\eta\chi}{\sqrt{1+\wt{v}v}}\beta_{1/2}^{(z)}\right)\nonumber\\
\phantom{\qquad=}{}\times\left(b_{-1}^{(z)}+\wt{b}_1^{(z)}-\frac{ 2\eta\chi \wt{v}}{\sqrt{1+\wt{v}v}}\beta_{-1/2}^{(z)}\right)\delta\left(-2\beta_{-1/2}^{(z)}+\frac{\eta\chi}{\sqrt{1+\tilde{v}v}}b_{-1}^{(z)}\right),\label{eq:BBd(B)}
\end{gather}
where \smash{${\rm e}^{-I_{\tilde{v}v\chi}}$} encodes the entire dependence of the action on the supermoduli $\wt{v};\!v|\chi$. We can determine this by Taylor series expansion in $\wt{v};\!v|\chi$ around $0;\!0|0$ and taking into account that under a generic change in supercomplex structure the action changes by an amount~$\delta I$ as displayed in~\eqref{eq:deltaI_cs}. The derivative of the action with respect to a supermodulus~$t$ is in turn given by \eqref{eq:partialIt}, and so we can completely reconstruct the quantity ${\rm e}^{-I_{\tilde{v}v\chi}}$ using this information.

Equation \eqref{eq:BBd(B)} is the main result of the current note, but to show that it leads to a sensible path integral it is necessary to also show that BRST-exact states decouple -- we discuss this next.

\section{Gauge invariance}\label{sec:GI}
One of the most important consistency checks of \eqref{eq:BBd(B)} is to show that when we insert a BRST-exact vertex operator into the path integral the latter should vanish, at least up to a total derivative in supermoduli space. The relevant point here is therefore that the insertion \eqref{eq:BBd(B)}, in particular, (anti-)commutes with the BRST charge up to a total derivative.

We will find it convenient in this section to work in terms of a different set of supermoduli that we will label by $\wt{\bm{z}};\!\bm{z}|\bm{\theta}$. In fact, we will only define these implicitly, via their variations, but this will be all we need for the purposes of this section. Let us primarily consider the superconformal vector fields \eqref{eq:VtVcomp}, in particular
\begin{gather}
\delta\wt{\V}(\wt{z})=\delta\wt{v}\wt{\V}_{\tilde{v}}(\wt{z})+\delta v\wt{\V}_v(\wt{z})+\delta\chi \wt{\V}_{\chi} (\wt{z}),\nonumber\\
\delta\V(z|\theta)=\delta\wt{v}\V_{\tilde{v}}(z|\theta)+\delta v\V_v(z|\theta)+\delta\chi \V_{\chi} (z|\theta),\label{eq:deltaVtdeltaVvtvchi}
\end{gather}
which according to \eqref{eq:VtVcomp} take the explicit form
\begin{gather}
\delta\wt{\V}=-\frac{\delta\wt{v}}{1+\wt{v}v}-\frac{\delta v}{1+\wt{v}v}\wt{z}^2,\nonumber\\
\delta\V
= -\frac{\delta\wt{v}}{1+\wt{v}v}\left(z^2-\frac{2\eta\chi}{\sqrt{1+\wt{v}v}}z\theta\right)
-\frac{\delta v}{1+\wt{v}v}\left(1+\frac{2\eta\chi\wt{v}}{\sqrt{1+\wt{v}v}}\theta\right)\nonumber\\
\phantom{\delta\V
=}{}+\frac{\eta\delta\chi}{\sqrt{1+\wt{v}v}}
\left(2\theta+\frac{\eta\chi}{\sqrt{1+\wt{v}v}}\right).\label{eq:deltaVtdeltaV}
\end{gather}
We can then extract the variations $\delta z(z|\theta)$ and $\delta\theta(z|\theta)$ from $\delta\V(z|\theta)$ using the identity
\begin{gather*}
\delta z = \delta\V-\frac{1}{2}\theta D_\theta \delta\V,\qquad
\delta\theta=\frac{1}{2}D_\theta \delta\V,
\end{gather*}
which in turn follow from the linearised superconformal condition $D_\theta \delta z=\theta D_\theta \delta \theta+\delta \theta$. In terms of these, we then have
\begin{gather}
\delta\wt{z}(\wt{z})=-\frac{\delta\wt{v}}{1+\wt{v}v}-\frac{\delta v}{1+\wt{v}v}\wt{z}^2,\nonumber\\
\delta z(z|\theta)=
-\frac{\delta\wt{v}}{1+\wt{v}v}\left(z^2-\frac{\eta\chi}{\sqrt{1+\wt{v}v}}z\theta\right)
-\frac{\delta v}{1+\wt{v}v}\left(1+\frac{\eta\chi\wt{v}}{\sqrt{1+\wt{v}v}}\theta\right)\nonumber\\
\phantom{\delta z(z|\theta)=}{}+\frac{\eta\delta\chi}{\sqrt{1+\wt{v}v}}
\left(\theta+\frac{\eta\chi}{\sqrt{1+\wt{v}v}}\right),\nonumber\\
\delta\theta(z|\theta)=
-\frac{\delta\wt{v}}{1+\wt{v}v}\left(z\theta+\frac{\eta\chi}{\sqrt{1+\wt{v}v}}z\right)
+\frac{\delta v}{1+\wt{v}v}\frac{\eta\chi\wt{v}}{\sqrt{1+\wt{v}v}}
-\frac{\eta\delta\chi}{\sqrt{1+\wt{v}v}},\label{eq:deltaztztheta}
\end{gather}
and we define the supermoduli variations $\delta\wt{\bm{z}};\!\delta\bm{z}|\delta\bm{\theta}$ in terms of these as the change in frame at the location of the puncture
\begin{gather}\label{eq:deltabfztztheta}
\delta\wt{\bm{z}}\dfn -\delta\wt{z}(0),\qquad
\delta\bm{z}\dfn -\delta z(0|0),\qquad
\delta\bm{\theta}\dfn -\delta\theta(0|0).
\end{gather}
From \eqref{eq:deltaztztheta} and \eqref{eq:deltabfztztheta}, it is seen that
\begin{gather}\label{eq:dztzthebf}
\delta \wt{\bm{z}} = \frac{\delta\wt{v}}{1+\wt{v}v},\qquad
\delta\bm{z}= \frac{\delta v}{1+\wt{v}v}-\frac{\delta\chi \chi}{1+\wt{v}v},\qquad
\delta\bm{\theta}= -\frac{\delta v\eta\chi\wt{v}}{(1+\wt{v}v)^{3/2}}+\frac{\delta\chi\eta}{\sqrt{1+\wt{v}v}}.
\end{gather}
We then rearrange the latter two relations and substitute the resulting set into \eqref{eq:deltaVtdeltaV} to extract expressions for $\delta\wt{\V}$ and $\delta\V$ in terms of the variations, $\delta\wt{\bm{z}};\!\delta\bm{z}|\delta\bm{\theta}$. By analogy to \eqref{eq:deltaVtdeltaVvtvchi}, we can write
\begin{gather*}
\delta\wt{\V}(\wt{z})=\delta\wt{\bm{z}}\wt{\V}_{\tilde{\bm{z}}}(\wt{z})+\delta \bm{z}\wt{\V}_{\bm{z}}(\wt{z})+\delta\bm{\theta} \wt{\V}_{\bm{\theta}} (\wt{z}),\\
\delta\V(z|\theta)=\delta\wt{\bm{z}}\V_{\tilde{\bm{z}}}(z|\theta)+\delta \bm{z}\V_{\bm{z}}(z|\theta)+\delta\bm{\theta} \V_{\bm{\theta}} (z|\theta),
\end{gather*}
and, in particular, this procedure leads to
\begin{gather}
\delta\wt{\V}=\delta\wt{\bm{z}}(-1)+\delta\bm{z}
\left(-\frac{1}{2}D_\theta\EuScript{R}_{\tilde{z}\theta}\wt{z}^2\right)+\delta\bm{\theta}\left(-\frac{1}{2}\EuScript{R}_{\tilde{z}\theta}\wt{z}^2\right),\nonumber\\
\delta\V=\delta{\wt{\bm{z}}}\left(-\frac{1}{2}D_\theta\EuScript{R}_{\tilde{z}\theta}z^2+\EuScript{R}_{\tilde{z}\theta}z\theta\right)+\delta\bm{z}(-1)+\delta\bm{\theta}(2\theta),\label{eq:deltaVtdeltaVdztdzdthbf}
\end{gather}
where we also took into account the super curvature expressions \eqref{eq:supercurv}, namely
\begin{gather}\label{eq:supercurv2}
\EuScript{R}_{\tilde{z}\theta} = \frac{2\eta\chi}{\sqrt{1+\wt{v}v}}\qquad{\rm and}\qquad
\frac{1}{2}D_\theta\EuScript{R}_{\tilde{z}\theta}=1.
\end{gather}
Of course, in the terms involving the combination $\frac{1}{2}D_\theta\EuScript{R}_{\tilde{z}\theta}$ we can trivially replace this\footnote{In fact, displaying the quantity $\frac{1}{2}D_\theta\EuScript{R}_{\tilde{z}\theta}$ explicitly, as we have done in \eqref{eq:deltaVtdeltaVdztdzdthbf}, requires that one carries out the computation with general super curvature before specifying that the underlying super Riemann surface is a super sphere, but (in order to keep the paper to a manageable size) I will not present the general super Riemann surface discussion here.} combination with 1, but we restored this explicitly to emphasise that the terms it appears multiplied by would have been absent in flat superspace (where instead $D_\theta\EuScript{R}_{\tilde{z}\theta}$ would equal zero). In particular, in conjunction with the first relation in \eqref{eq:supercurv2}, it allows us to differentiate between terms that appear due to super curvature of the super Riemann surface and terms that would have been present also in the absence of super curvature.

To emphasise this point, according to the second relation in \eqref{eq:supercurv2}, it would be clearly inconsistent to naively project onto flat superspace by setting $\R_{\tilde{z}\theta}=D_\theta\EuScript{R}_{\tilde{z}\theta}=0$, which is what we would have arrived at had we assumed a holomorphic splitting for supermoduli space. The presence of super curvature is mixing chiral and anti-chiral contributions in~\eqref{eq:deltaVtdeltaVdztdzdthbf} in a manner that cannot be removed by a change of coordinates. Indeed, there is a topological obstruction to a ``good'' holomorphic splitting~\cite{Nelson88b} on a super Riemann surface with the topology of a 2-sphere.

Given the superconformal vector fields \eqref{eq:deltaVtdeltaVdztdzdthbf}, the corresponding superghost contributions to the measure analogous to \eqref{eq:NSpuncmeasure} in terms of $\wt{\bm{z}};\!\bm{z}|\bm{\theta}$ take the form
\begin{gather}\label{eq:NSpuncmeasure2}
\int \mathcal{D}(\wt{\bm{z}},\!\bm{z}|\bm{\theta}){\rm e}^{-I_{\tilde{\bm{z}}\bm{z}\bm{\theta}}}\delta\big(\h{\EuScript{B}}_{\tilde{\bm{z}}}\big)\delta\big(\h{\EuScript{B}}_{\bm{z}}\big)\delta\big(\h{\EuScript{B}}_{\bm{\theta}}\big),
\end{gather}
where, as always, we adopt the shorthand $\mathcal{D}(\wt{\bm{z}},\!\bm{z}|\bm{\theta})=-{\rm i}[\rmd \wt{\bm{z}},\!\rmd \bm{z}|\rmd\bm{\theta}]$ and the quantity $I_{\tilde{\bm{z}}\bm{z}\bm{\theta}}$ encodes the entire supermoduli dependence of the full worldsheet superconformal field theory action. We can actually think of this as the full (matter plus ghosts) heterotic string theory action~\eqref{eq:fullaction.z}, discussed in further detail below. We will write
$
I_{\tilde{\bm{z}}\bm{z}\bm{\theta}}=I_{\tilde{\bm{z}}\bm{z}0}+\bm{\theta}\partial_{\bm{\theta}}I_{\tilde{\bm{z}}\bm{z}0}=I_{\tilde{\bm{z}}\bm{z}0}-\bm{\theta}\h{\partial}_{\bm{\theta}}$,
to denote the corresponding Taylor series expansion in $\bm{\theta}$. We have taken into account the relation~\eqref{eq:partialIt}, see also \eqref{eq:hDcommutators}.
Mapping $[\rmd \wt{\bm{z}},\!\rmd \bm{z}|\rmd\bm{\theta}]$ to the integral form $\rmd \wt{\bm{z}} \rmd \bm{z} \delta(\rmd\bm{\theta})$ (where each of the terms $\rmd \wt{\bm{z}}$, $\rmd \bm{z}$ and $\delta(\rmd\bm{\theta})$ have Grassmann-odd parity), making use of \eqref{eq:dztzthebf} and expanding the delta function, it easily follows that the supermoduli measures are related as follows
$
\rmd \wt{\bm{z}} \rmd \bm{z} \delta(\rmd\bm{\theta}) = \rmd \wt{v} \rmd v \delta(\rmd\chi) \eta(1+\wt{v}v)^{-3/2}$.
(This is of course the same as the conclusion reached in \eqref{eq:DDBer}, but we included this alternative derivation for variety.)
So the overall factors outside the parentheses in the superghost expressions \eqref{eq:BBB} are absorbed into the measure $\mathcal{D}(\wt{\bm{z}},\!\bm{z}|\bm{\theta})$ in the parametrisation in \eqref{eq:NSpuncmeasure2} that is determined by the superconformal vector fields \eqref{eq:deltaVtdeltaVdztdzdthbf}. The derivation of \eqref{eq:BBB} from \eqref{eq:Bthat3} is in turn precisely analogous to the corresponding derivation leading to the following superghost contributions to the measure:
\begin{gather}
\h{\EuScript{B}}_{\tilde{\bm{z}}}
=\wt{b}_{-1}+\frac{1}{2}D_\theta\EuScript{R}_{\tilde{z}\theta}b_1+\EuScript{R}_{\tilde{z}\theta} \beta_{1/2},\qquad
\h{\EuScript{B}}_{\bm{z}}
=b_{-1}+\frac{1}{2}D_\theta\EuScript{R}_{\tilde{z}\theta}\wt{b}_1,\nonumber\\
\h{\EuScript{B}}_{\bm{\theta}}
=-2\beta_{-1/2}-\frac{1}{2}\EuScript{R}_{\tilde{z}\theta}\wt{b}_{1},\qquad
\h{\EuScript{B}}_{\bm{\theta}}+\bm{\theta}\h{\EuScript{B}}_{\bm{z}}
\equiv -2\beta_{-1/2}+\bm{\theta}b_{-1}-\EuScript{K}\wt{b}_1,\label{eq:BBB2x}
\end{gather}
where instead of \eqref{eq:VtVcomp} that led to \eqref{eq:BBB} in the previous derivation we now made use of \eqref{eq:deltaVtdeltaVdztdzdthbf} to arrive at \eqref{eq:BBB2x}. We have also defined the parity-odd quantity
\begin{gather}\label{eq:K}
\EuScript{K}\dfn \frac{1}{2}\big(\EuScript{R}_{\tilde{z}\theta}-\bm{\theta}D_\theta\EuScript{R}_{\tilde{z}\theta}\big).
\end{gather}
It will turn out that gauge invariance requires $\EuScript{K}=0$, but we want to remain agnostic about the precise coefficient of proportionality relating $\EuScript{R}_{\tilde{z}\theta}$ to $\bm{\theta}$ at this point. It is useful to adopt specific notation for the corresponding BRST (anti-)commutators
\begin{gather}
\h{\partial}_{\tilde{\bm{z}}}= \big\{Q_B,\h{\EuScript{B}}_{\tilde{\bm{z}}}\big\},\qquad
\h{\partial}_{\bm{z}}=\big\{Q_B,\h{\EuScript{B}}_{\bm{z}}\big\},\qquad
\h{D}_{\bm{\theta}}=\big[Q_B,\h{\EuScript{B}}_{\bm{\theta}}\big],\nonumber
\\
\h{\partial}_{\bm{\theta}}=\big[Q_B,\h{\EuScript{B}}_{\bm{\theta}}+\bm{\theta}\h{\EuScript{B}}_{\bm{z}}\big],\label{eq:QBBcomms}
\end{gather}
where, in particular,
\begin{gather}
\h{\partial}_{\tilde{\bm{z}}}
 =\wt{L}_{-1}+\frac{1}{2}D_{\theta}\EuScript{R}_{\tilde{z}\theta} L_1+\frac{1}{2}\EuScript{R}_{\tilde{z}\theta} G_{1/2},\qquad
\h{\partial}_{\bm{z}}
 = L_{-1}+ \frac{1}{2}D_{\theta}\EuScript{R}_{\tilde{z}\theta} \wt{L}_1,\nonumber\\
\h{D}_{\bm\theta}
 = G_{-1/2}+\frac{1}{2}\EuScript{R}_{\tilde{z}\theta} \wt{L}_1,\qquad
\h{\partial}_{\bm{\theta}}
=G_{-1/2}-\bm{\theta}L_{-1}+\EuScript{K}\wt{L}_1,\label{eq:hDcommutators}
\end{gather}
which follow from the defining relation \eqref{eq:QBBcomms} and explicit evaluation of the various (anti-)com\-mu\-tators.

The operators in \eqref{eq:hDcommutators} are very much like derivative operators, but the objects they act on must not be annihilated by the various super Virasoro generators appearing in \eqref{eq:hDcommutators} in order to give a non-vanishing answer. So these operators do not quite replace the ordinary notion of a derivative. Clearly, there will also be local functions or superconformal tensors (such as $\EuScript{R}_{\tilde{z}\theta}$) that have non-trivial supermoduli dependence while nevertheless being annihilated by $\h{\partial}_t$. So to complete the story we need to add ordinary supermoduli derivatives to the right-hand sides in~\eqref{eq:hDcommutators} in order to be able to properly identify them with derivative operators that can act on both operators and ordinary superconformal tensors or functions (that do not necessarily have any operator dependence). Therefore, the total derivatives arising from the BRST \mbox{(anti-)com}\-mu\-tators that we expect to find should be of the form $\partial_t+\h{\partial}_t$. (We will see in \eqref{eq:BBBdBn2} and~\eqref{eq:hpartialtot} that this is the combination that arises naturally.) However, as we briefly summarise momentarily, the $\h{\partial}_t$ contributions can be replaced by ordinary $\partial_t$ derivatives of the worldsheet action (the precise relation being $\partial_t {\rm e}^{-I} = {\rm e}^{-I}\h{\partial}_t$, where $I$ is the full worldsheet action of interest, namely that of the heterotic string plus superghosts
\begin{gather}\label{eq:fullaction.z}
I = \frac{1}{2\pi} \int\mathcal{D}(\wt{z},\!z|\theta)\left(\frac{1}{\alpha'}\partial_{\tilde{z}}X D_\theta X+\Lambda  D_\theta \Lambda+B_{z\theta}\partial_{\tilde{z}}C^z-\wt{B}_{\tilde{z}\tilde{z}}D_\theta \wt{C}^{\tilde{z}}\right),
\end{gather}
where we will set $\alpha'=2$ and (apart from the minus sign in the last term in \eqref{eq:fullaction.z} which turns out to be important, e.g., in the derivation of \eqref{eq:deltaI_cs}) we have adopted the notation in~\cite{Witten12b}. Briefly,
writing $I=I_{\rm matter}+I_{\rm ghosts}$, the matter sector $I_{\rm matter}$ receives contributions from the scalar superfields $X^{\mu}(\wt{z};\!z|\theta)$, $\mu=0,\dots,9$, that map the string worldsheet $\Sigma$ into flat Euclidean spacetime $\mathbf{R}^{10}$, and the current algebra fermions $\Lambda_a(\wt{z})$ with $a=1,\dots,32$. The latter correspond to spinor superfields taking values in $\mathit{\Pi}\EuScript{L}$, i.e., they are fermionic fields taking values in a square root $\EuScript{L}$ of the line bundle ${\Ber}(\Sigma_L)$. The argument of $\Lambda_a(\wt{z})$ is meant to indicate that the line bundle $\EuScript{L}$ is anti-holomorphic, so that it can be constructed using anti-holomorphic transition functions (that in the indicated chart are functions of $\wt{z}$ only) and so commute with~$D_\theta$. Accordingly, the $\Lambda$-matter sector in~\eqref{eq:fullaction.z} is a section of $\EuScript{L}^2\otimes \mathcal{D}^{-1}\cong {\Ber}(\Sigma_L\times \Sigma_R)$ and can therefore be integrated. See \cite[Sections~3.1--3.3]{Witten12b} for further detail. Sums over $\mu$ and $a$ are implicit in \eqref{eq:fullaction.z}. The superghost sector $I_{\rm ghosts}$ of the action receives contributions from the superghosts $B(z|\theta)$ and $C(z|\theta)$, which are sections of $\mathcal{D}^{-3}$ and $\mathit{\Pi}\mathcal{D}^2$, respectively, and the anti-chiral superghosts $\wt{B}(\wt{z})$ and $\wt{C}(\wt{z})$, which are sections of $\mathit{\Pi}{\Ber}(\Sigma_L)^2$ and $\mathit{\Pi}{\Ber}(\Sigma_L)^{-1}$, respectively.\looseness=-1

The conclusion that $\partial_t {\rm e}^{-I} = {\rm e}^{-I}\h{\partial}_t$ is then derived as follows.
We parametrise a small change in superconformal structure as a change in superfields, $X,\Lambda,B,\dots$, generated by locally-defined quasi-superconformal vector superfields $\delta\wt{\mathcal{V}}^{\tilde{z}}$, $\delta\mathcal{V}^{z}$ keeping the worldsheet superconformal frame fixed,
\begin{gather}
\delta X = -\delta\wt{\mathcal{V}}^{\tilde{z}}\partial_{\tilde{z}}X-\delta\mathcal{V}^{z}\partial_zX-\tfrac{1}{2}D_\theta \delta\mathcal{V}^{z}D_\theta X,\qquad
\delta\Lambda  = - \delta\wt{\mathcal{V}}^{\tilde{z}}\partial_{\tilde{z}}\Lambda-\tfrac{1}{2}\partial_{\tilde{z}}\delta\wt{\mathcal{V}}^{\tilde{z}}\Lambda,\nonumber\\
\delta B_{z\theta} =-\delta\mathcal{V}^{z}\partial_zB_{z\theta}-\tfrac{1}{2}D_\theta \delta\mathcal{V}^{z}D_\theta B_{z\theta} - \tfrac{3}{2}\partial_z\delta\mathcal{V}^{z}B_{z\theta},\nonumber\\
\delta C^z =-\delta\mathcal{V}^{z}\partial_zC^z-\tfrac{1}{2}D_\theta\delta \mathcal{V}^{z}D_\theta C^z +\partial_z\delta\mathcal{V}^{z}C^z,\nonumber\\
\delta \wt{B}_{\tilde{z}\tilde{z}} =- \delta\wt{\mathcal{V}}^{\tilde{z}}\partial_{\tilde{z}}\wt{B}_{\tilde{z}\tilde{z}}-2\partial_{\tilde{z}}\delta\wt{\mathcal{V}}^{\tilde{z}}\wt{B}_{\tilde{z}\tilde{z}},\qquad
\delta \wt{C}^{\tilde{z}} =- \delta\wt{\mathcal{V}}^{\tilde{z}}\partial_{\tilde{z}}\wt{C}^{\tilde{z}}+\partial_{\tilde{z}}\delta\wt{\mathcal{V}}^{\tilde{z}}\wt{C}^{\tilde{z}}.\label{eq:deltaXBCetc-scmain}
\end{gather}
These variations are essentially super Lie derivatives \cite{VerlindeHphd}, the precise expressions follow from knowledge of the spaces in which these superfields take their values \cite{Witten12b}.
The corresponding change in the action induced by
\eqref{eq:deltaXBCetc-scmain} is given by
\begin{gather}
\delta I = \frac{1}{2\pi}\int\mathcal{D}(\wt{z},z|\theta)\big[(\partial_{\tilde{z}}\delta\mathcal{V}^z)\mathcal{S}_{z\theta}+\big(D_\theta \delta\wt{\mathcal{V}}^{\tilde{z}}\big)\wt{T}_{\tilde{z}\tilde{z}}\big].\label{eq:deltaI_cs}
\end{gather}
The chiral and anti-chiral halves of the total energy-momentum tensors are defined by $\mathcal{S}_{z\theta}=\mathcal{S}_X+\mathcal{S}_{BC}$ and $\wt{T}_{\tilde{z}\tilde{z}}=\wt{T}_X+\wt{T}_{\Lambda}+\wt{T}_{\tilde{B}\tilde{C}}$, respectively, where the various contributions are, in turn, found to take the standard form
\begin{gather*}
\mathcal{S}_X = -\frac{1}{\alpha'}D_\theta X D_\theta^2X,\qquad
\mathcal{S}_{BC}=\frac{1}{2}D_\theta B_{z\theta}D_\theta C^z-\frac{3}{2}D_\theta^2C^zB_{z\theta}-C^zD_\theta^2 B_{z\theta}\\
\wt{T}_X=-\frac{1}{\alpha'}\partial_{\tilde{z}}X \partial_{\tilde{z}}X,\qquad
\wt{T}_\Lambda =-\partial_{\tilde{z}}\Lambda \Lambda,\qquad
\wt{T}_{\tilde{B}\tilde{C}}=2\partial_{\tilde{z}}\wt{C}^{\tilde{z}}\wt{B}_{\tilde{z}\tilde{z}}+\wt{C}^{\tilde{z}}\partial_{\tilde{z}}\wt{B}_{\tilde{z}\tilde{z}}.
\end{gather*}
It is instructive to compare this to the component formulation given in \cite{D'HokerPhong15b}.
Integrating by parts in \eqref{eq:deltaI_cs} actually produces a boundary term. Cancelling this by an appropriate addition to the action, and using the ordinary chain rule to map the \emph{total} change $\delta \mathcal{V}$, $\delta\wt{\mathcal{V}}$ (which are smooth in $\wt{z};\!z|\theta$) to a change in $\delta\V$, $\delta\wt{\V}$ keeping coordinates $\wt{w};\!w|\psi$ fixed (which as we have shown is superconformal in $z|\theta$ and $\wt{z}$ respectively and given for our current purposes by \eqref{eq:deltaVtdeltaVdztdzdthbf}), one arrives at the following result for the derivative of the action with respect to a change in a~supermodulus~$t$,
\begin{gather}\label{eq:partialIt}
\frac{\partial I}{\partial t} = - \h{\partial}_t
\end{gather}
with $\h{\partial}_t$ as given (in the case of interest) in \eqref{eq:hDcommutators}. We have assumed that the only chart overlap here is $\U_w\cap \U_z$, which is the case of interest when moving NS punctures across the super Riemann surface. (We are again omitting details here.) So we conclude that indeed $\partial_t {\rm e}^{-I} = {\rm e}^{-I}\h{\partial}_t$ as advertised above. Ultimately therefore, the total derivatives associated to the decoupling of BRST-exact vertex operators will be entirely constructed out of ordinary $\partial_t$ derivatives.\looseness=1

We are now well-equipped to consider the BRST (anti-)commutator associated to the decoupling of a BRST-exact vertex operator insertion in the presence of an NS puncture, whose location on the underlying super Riemann surface is associated to a supermodulus of even$|$odd dimension $2|1$. When we insert a BRST-exact vertex operator into the path integral we expect this to decouple. This is in turn required to preserve gauge invariance. In demonstrating this decoupling one encounters a number of (anti-)commutators as one unwraps the BRST charge contour off the said BRST-exact vertex operator. In the absence of other vertex operator insertions or supermoduli contributions to the measure this is trivially zero, because there is no obstruction to unwrapping the contour to a point, whereby it can be seen to vanish since the $\mathrm{OSp}(2,1)$ vacuum (represented by the unit operator insertion) is annihilated by the BRST charge (the BRST current has non-singular OPE with the unit operator and can hence be Taylor expanded around $z|\theta=0|0$). If however there are supermoduli present (which might be associated to handle supermoduli or other external vertex operators) then the BRST charge contour encounters superghost contributions associated to the gauge slice of our choice. The latter is determined by how we parametrise the integral over supermoduli. The terms of interest are the superghost contributions associated to translating a NS puncture across the super Riemann surface. If the underlying super Riemann surface has the topology of a 2-sphere, the said measure contributions are given by~\eqref{eq:NSpuncmeasure2}, where in particular these take the form displayed in~\eqref{eq:BBB2x}. On a~more general super Riemann surface the insertions are similar but in general contain additional terms associated to higher derivatives of super curvature. So the (anti-)commutator that we encounter as we try to unwrap the BRST charge off the surface is the following
$
\big\{Q_B,\h{\EuScript{B}}_{\tilde{\bm{z}}}\h{\EuScript{B}}_{\bm{z}}\delta\big(\h{\EuScript{B}}_{\bm{\theta}}+\bm{\theta}\h{\EuScript{B}}_{\bm{z}}\big)\big]$.
We have inserted an additional factor of \smash{$\bm{\theta}\h{\EuScript{B}}_{\bm{z}}$} in the argument of the delta function. It will turn out to be convenient to do so, but the point to notice is that it is equivalent to the original insertion since $\h{\EuScript{B}}_{\bm{z}}$ is Grassmann-odd. Denoting the Grassmann parity of $\delta\big(\h{\EuScript{B}}_{\bm{\theta}}+\bm{\theta}\h{\EuScript{B}}_{\bm{z}}\big)$ by $|\delta|$, and taking into account that the Grassmann parities of the BRST charge $Q_B$ and that of the insertions $\h{\EuScript{B}}_{\tilde{\bm{z}}}$ and $\h{\EuScript{B}}_{\bm{z}}$ are odd, and that the parity of $\h{\EuScript{B}}_{\bm{\theta}}$ is even, it immediately follows that
\begin{gather}
\big\{Q_B,\h{\EuScript{B}}_{\tilde{\bm{z}}}\h{\EuScript{B}}_{\bm{z}}\delta\big(\h{\EuScript{B}}_{\bm{\theta}}+\bm{\theta}\h{\EuScript{B}}_{\bm{z}}\big)\big]\nonumber\\
\qquad=
\big\{Q_B,\h{\EuScript{B}}_{\tilde{\bm{z}}}\big\}\h{\EuScript{B}}_{\bm{z}}\delta\big(\h{\EuScript{B}}_{\bm{\theta}}+\bm{\theta}\h{\EuScript{B}}_{\bm{z}}\big)
-\big\{Q_B,\h{\EuScript{B}}_{\bm{z}}\big\}\h{\EuScript{B}}_{\tilde{\bm{z}}}\delta\big(\h{\EuScript{B}}_{\bm{\theta}}+\bm{\theta}\h{\EuScript{B}}_{\bm{z}}\big)
+\big\{Q_B,\delta\big(\h{\EuScript{B}}_{\bm{\theta}}+\bm{\theta}\h{\EuScript{B}}_{\bm{z}}\big)\big]\h{\EuScript{B}}_{\tilde{\bm{z}}}\h{\EuScript{B}}_{\bm{z}}\nonumber\\
\phantom{\qquad=}{}-\big[\h{\EuScript{B}}_{\tilde{\bm{z}}},\big\{Q_B,\h{\EuScript{B}}_{\bm{z}}\big\}\big]\delta\big(\h{\EuScript{B}}_{\bm{\theta}}
+\bm{\theta}\h{\EuScript{B}}_{\bm{z}}\big)
-(-)^{|\delta|}
\big\{\h{\EuScript{B}}_{\tilde{\bm{z}}},\big\{Q_B,\delta\big(\h{\EuScript{B}}_{\bm{\theta}}+\bm{\theta}\h{\EuScript{B}}_{\bm{z}}\big)\big]\big]\h{\EuScript{B}}_{\bm{z}}\nonumber\\
\phantom{\qquad=}{}+(-)^{|\delta|}\big\{\h{\EuScript{B}}_{\bm{z}},\big\{Q_B,\delta\big(\h{\EuScript{B}}_{\bm{\theta}}+\bm{\theta}\h{\EuScript{B}}_{\bm{z}}\big)\big]\big]\h{\EuScript{B}}_{\tilde{\bm{z}}}
+\big\{\h{\EuScript{B}}_{\tilde{\bm{z}}},\big\{\h{\EuScript{B}}_{\bm{z}},\big\{Q_B,\delta\big(\h{\EuScript{B}}_{\bm{\theta}}
+\bm{\theta}\h{\EuScript{B}}_{\bm{z}}\big)\big]\big]\big].\label{eq:BBBdBn}
\end{gather}
The last term vanishes, ultimately, due to the fact that the $\big\{\h{\EuScript{B}}_s,\h{\EuScript{B}}_t\big]=0$. Let us rewrite \eqref{eq:BBBdBn} in terms of the derivative operators defined in \eqref{eq:QBBcomms},
\begin{gather}
\big\{Q_B,\h{\EuScript{B}}_{\tilde{\bm{z}}}\h{\EuScript{B}}_{\bm{z}}\delta\big(\h{\EuScript{B}}_{\bm{\theta}}+\bm{\theta}\h{\EuScript{B}}_{\bm{z}}\big)\big]\nonumber\\
\qquad=
 \h{\partial}_{\tilde{\bm{z}}}\h{\EuScript{B}}_{\bm{z}}\delta\big(\h{\EuScript{B}}_{\bm{\theta}}+\bm{\theta}\h{\EuScript{B}}_{\bm{z}}\big)
- \h{\partial}_{\bm{z}}\h{\EuScript{B}}_{\tilde{\bm{z}}}\delta\big(\h{\EuScript{B}}_{\bm{\theta}}+\bm{\theta}\h{\EuScript{B}}_{\bm{z}}\big)
+\big\{Q_B,\delta\big(\h{\EuScript{B}}_{\bm{\theta}}+\bm{\theta}\h{\EuScript{B}}_{\bm{z}}\big)\big]\h{\EuScript{B}}_{\tilde{\bm{z}}}\h{\EuScript{B}}_{\bm{z}}\nonumber\\
\phantom{\qquad=}{}-\big[\h{\EuScript{B}}_{\tilde{\bm{z}}},\big\{Q_B,\h{\EuScript{B}}_{\bm{z}}\big\}\big]\delta\big(\h{\EuScript{B}}_{\bm{\theta}}+\bm{\theta}\h{\EuScript{B}}_{\bm{z}}\big)
-(-)^{|\delta|}
\big\{\h{\EuScript{B}}_{\tilde{\bm{z}}},\big\{Q_B,\delta\big(\h{\EuScript{B}}_{\bm{\theta}}+\bm{\theta}\h{\EuScript{B}}_{\bm{z}}\big)\big]\big]\h{\EuScript{B}}_{\bm{z}}\nonumber\\
\phantom{\qquad=}{}+(-)^{|\delta|}\big\{\h{\EuScript{B}}_{\bm{z}},\big\{Q_B,\delta\big(\h{\EuScript{B}}_{\bm{\theta}}
+\bm{\theta}\h{\EuScript{B}}_{\bm{z}}\big)\big]\big]\h{\EuScript{B}}_{\tilde{\bm{z}}}.\label{eq:BBBdBn2x}
\end{gather}
For the commutators involving $\delta\big(\h{\EuScript{B}}_{\bm{\theta}}+\bm{\theta}\h{\EuScript{B}}_{\bm{z}}\big)$, we also need the following identity. Since $\big[\h{\EuScript{B}}_{t_1},\big[\h{\EuScript{B}}_{t_2},\allowbreak\smash{\big[Q_B,\h{\EuScript{B}}_{t_3}\big]\big]\big] =0}$, for any set of supermoduli $t_j$ it is not too hard to show that
\begin{align*}
\big\{Q_B,\delta\big(\h{\EuScript{B}}_{\bm{\theta}}+\bm{\theta}\h{\EuScript{B}}_{\bm{z}}\big)\big] = {}& \big[Q_B,\h{\EuScript{B}}_{\bm{\theta}}+\bm{\theta}\h{\EuScript{B}}_{\bm{z}}\big] \delta'\big(\h{\EuScript{B}}_{\bm{\theta}}+\bm{\theta}\h{\EuScript{B}}_{\bm{z}}\big)\\
&+\frac{1}{2} \big[\h{\EuScript{B}}_{\bm{\theta}}+\bm{\theta}\h{\EuScript{B}}_{\bm{z}},\big[Q_B,\h{\EuScript{B}}_{\bm{\theta}}+\bm{\theta}\h{\EuScript{B}}_{\bm{z}}\big]\big] \delta''\big(\h{\EuScript{B}}_{\bm{\theta}}+\bm{\theta}\h{\EuScript{B}}_{\bm{z}}\big).
\end{align*}
This follows from pure combinatorics (rather than any detailed properties of these operators). From \eqref{eq:QBBcomms} and also
\begin{gather*}
\frac{1}{2} \big[\h{\EuScript{B}}_{\bm{\theta}}+\bm{\theta}\h{\EuScript{B}}_{\bm{z}},\big[Q_B,\h{\EuScript{B}}_{\bm{\theta}}+\bm{\theta}\h{\EuScript{B}}_{\bm{z}}\big]\big]
=b_{-1}=\partial_{\bm{\theta}}\big(\h{\EuScript{B}}_{\bm{\theta}}+\bm{\theta}\h{\EuScript{B}}_{\bm{z}}\big)+(\partial_{\bm{\theta}}\EuScript{K}) \wt{b}_1,
\end{gather*}
where we took into account the explicit expression for $\h{\EuScript{B}}_{\bm{\theta}}$ in \eqref{eq:BBB2x}, we learn that
\begin{align*}
\big\{Q_B,\delta\big(\h{\EuScript{B}}_{\bm{\theta}}+\bm{\theta}\h{\EuScript{B}}_{\bm{z}}\big)\big]
&= \h{\partial}_{\bm{\theta}} \delta'\big(\h{\EuScript{B}}_{\bm{\theta}}+\bm{\theta}\h{\EuScript{B}}_{\bm{z}}\big)+\big(\partial_{\bm{\theta}}\delta'
\big(\h{\EuScript{B}}_{\bm{\theta}}+\bm{\theta}\h{\EuScript{B}}_{\bm{z}}\big)\big)
+(\partial_{\bm{\theta}}\EuScript{K}) \wt{b}_1\delta''\big(\h{\EuScript{B}}_{\bm{\theta}}+\bm{\theta}\h{\EuScript{B}}_{\bm{z}}\big)\\
&= \h{\partial}_{\bm{\theta}} \delta'\big(\h{\EuScript{B}}_{\bm{\theta}}+\bm{\theta}\h{\EuScript{B}}_{\bm{z}}\big)+b_{-1}\delta''\big(\h{\EuScript{B}}_{\bm{\theta}}+\bm{\theta}\h{\EuScript{B}}_{\bm{z}}\big).
\end{align*}
Computing the remaining (anti-)commutators, we find
\begin{gather*}
-\big[\h{\EuScript{B}}_{\tilde{\bm{z}}},\big\{Q_B,\h{\EuScript{B}}_{\bm{z}}\big\}\big]\delta\big(\h{\EuScript{B}}_{\bm{\theta}}+\bm{\theta}\h{\EuScript{B}}_{\bm{z}}\big)
=-\big(D_\theta\EuScript{R}_{\tilde{z}\theta}\big(b_0-\wt{b}_0\big)
+\EuScript{R}_{\tilde{z}\theta}\beta_{-1/2}\big)\delta\big(\h{\EuScript{B}}_{\bm{\theta}}+\bm{\theta}\h{\EuScript{B}}_{\bm{z}}\big),
\\
-(-)^{|\delta|}
\big\{\h{\EuScript{B}}_{\tilde{\bm{z}}},\big\{Q_B,\delta\big(\h{\EuScript{B}}_{\bm{\theta}}+\bm{\theta}\h{\EuScript{B}}_{\bm{z}}\big)\big]\big]\h{\EuScript{B}}_{\bm{z}}\\
\qquad=
D_\theta\EuScript{R}_{\tilde{z}\theta}\beta_{1/2}\h{\EuScript{B}}_{\bm{z}}\delta'\big(\h{\EuScript{B}}_{\bm{\theta}}+\bm{\theta}\h{\EuScript{B}}_{\bm{z}}\big)
+2\EuScript{K}\big(b_{0}-\wt{b}_0+\bm{\theta}\beta_{-1/2}\big)\h{\EuScript{B}}_{\bm{z}}\delta'\big(\h{\EuScript{B}}_{\bm{\theta}} +\bm{\theta}\h{\EuScript{B}}_{\bm{z}}\big)
\end{gather*}
and
$
+(-)^{|\delta|}\big\{\h{\EuScript{B}}_{\bm{z}},\big\{Q_B,\delta\big(\h{\EuScript{B}}_{\bm{\theta}}+\bm{\theta}\h{\EuScript{B}}_{\bm{z}}\big)\big]\big]\h{\EuScript{B}}_{\tilde{\bm{z}}}=0$.
Collecting these results and substituting them into \eqref{eq:BBBdBn2x} implies that
\begin{gather}
\big\{Q_B,\h{\EuScript{B}}_{\tilde{\bm{z}}}\h{\EuScript{B}}_{\bm{z}}\delta\big(\h{\EuScript{B}}_{\bm{\theta}}+\bm{\theta}\h{\EuScript{B}}_{\bm{z}}\big)\big]\nonumber\\
\qquad= \h{\partial}_{\tilde{\bm{z}}}^{\rm total}\Big(\h{\EuScript{B}}_{\bm{z}}\delta\big(\h{\EuScript{B}}_{\bm{\theta}}+\bm{\theta}\h{\EuScript{B}}_{\bm{z}}\big)\big)
- \h{\partial}_{\bm{z}}^{\rm total}\big(\h{\EuScript{B}}_{\tilde{\bm{z}}}\delta\big(\h{\EuScript{B}}_{\bm{\theta}}+\bm{\theta}\h{\EuScript{B}}_{\bm{z}}\big)\big)
+\h{\partial}_{\bm{\theta}}^{\rm total}\big(\h{\EuScript{B}}_{\tilde{\bm{z}}}\h{\EuScript{B}}_{\bm{z}}\delta'\big(\h{\EuScript{B}}_{\bm{\theta}}+\bm{\theta}\h{\EuScript{B}}_{\bm{z}}\big)\big)\nonumber\\
\phantom{\qquad=}{}
-D_\theta\EuScript{R}_{\tilde{z}\theta}\big(b_0-\wt{b}_0\big)\delta\big(\h{\EuScript{B}}_{\bm{\theta}}+\bm{\theta}\h{\EuScript{B}}_{\bm{z}}\big)
-\EuScript{K}\bm{\theta}\h{\EuScript{B}}_{\bm{z}}\delta\big(\h{\EuScript{B}}_{\bm{\theta}}+\bm{\theta}\h{\EuScript{B}}_{\bm{z}}\big)\nonumber\\
\phantom{\qquad=}{}
+ \big(\partial_{\tilde{\bm{z}}}\EuScript{K} \wt{b}_1
\big)\h{\EuScript{B}}_{\bm{z}}\delta'\big(\h{\EuScript{B}}_{\bm{\theta}}+\bm{\theta}\h{\EuScript{B}}_{\bm{z}}\big)
-\big(\partial_{\bm{z}}\EuScript{K} \wt{b}_1\big)\h{\EuScript{B}}_{\tilde{\bm{z}}}\delta'\big(\h{\EuScript{B}}_{\bm{\theta}}+\bm{\theta}\h{\EuScript{B}}_{\bm{z}}\big)\nonumber\\
\phantom{\qquad=}{}+(\partial_{\bm{\theta}}\EuScript{K}) \wt{b}_1\h{\EuScript{B}}_{\tilde{\bm{z}}}\h{\EuScript{B}}_{\bm{z}}\delta''\big(\h{\EuScript{B}}_{\bm{\theta}}+\bm{\theta}\h{\EuScript{B}}_{\bm{z}}\big)
+2\EuScript{K}\big(b_{0}-\wt{b}_0+\bm{\theta}\beta_{-1/2}\big)\h{\EuScript{B}}_{\bm{z}}\delta'\big(\h{\EuScript{B}}_{\bm{\theta}}+\bm{\theta}\h{\EuScript{B}}_{\bm{z}}\big)\nonumber\\
 \phantom{\qquad=}{}+(-2\partial_{\bm{\theta}}\EuScript{K} \beta_{1/2})\h{\EuScript{B}}_{\bm{z}}\delta'\big(\h{\EuScript{B}}_{\bm{\theta}}+\bm{\theta}\h{\EuScript{B}}_{\bm{z}}\big)
+\big(2\partial_{\bm{z}}\EuScript{K} \beta_{1/2}\big)\delta\big(\h{\EuScript{B}}_{\bm{\theta}}+\bm{\theta}\h{\EuScript{B}}_{\bm{z}}\big).\label{eq:BBBdBn2}
\end{gather}
We have written
\begin{gather}\label{eq:hpartialtot}
\h{\partial}_t^{\rm total}=\h{\partial}_{t}+\partial_{t}
\end{gather}
and have made use of a number of relations in arriving at this result, all of which have been determined from the explicit representations given in \eqref{eq:BBB2x} and the definition of $\EuScript{K}$ in \eqref{eq:K}\looseness=-1
\begin{gather*}
\partial_{\tilde{\bm{z}}}\h{\EuScript{B}}_{\bm{z}}=0,\qquad
\partial_{\bm{z}}\big(\h{\EuScript{B}}_{\bm{\theta}}+\bm{\theta}\h{\EuScript{B}}_{\bm{z}}\big)= -\partial_{\bm{z}}\EuScript{K} \wt{b}_1,\qquad
\partial_{\bm{z}}\h{\EuScript{B}}_{\tilde{\bm{z}}}=\partial_{\bm{z}}\EuScript{R}_{\tilde{z}\theta} \beta_{1/2},\qquad
\partial_{\bm{\theta}}\h{\EuScript{B}}_{\tilde{\bm{z}}}=\partial_{\bm{\theta}}\EuScript{R}_{\tilde{z}\theta} \beta_{1/2},\\
\partial_{\bm{\theta}}\h{\EuScript{B}}_{\bm{z}}=0,\qquad
\partial_{\tilde{\bm{z}}}\big(\h{\EuScript{B}}_{\bm{\theta}}+\bm{\theta}\h{\EuScript{B}}_{\bm{z}}\big)=-\partial_{\tilde{\bm{z}}}\EuScript{K} \wt{b}_1,\qquad
\EuScript{R}_{\tilde{z}\theta}(\!-2\beta_{-1/2})
=\EuScript{R}_{\tilde{z}\theta}\big(\h{\EuScript{B}}_{\bm{\theta}}+\bm{\theta}\h{\EuScript{B}}_{\bm{z}}\big)
-2\EuScript{K}\bm{\theta}\h{\EuScript{B}}_{\bm{z}}.
\end{gather*}
Gauge invariance requires that only the total derivative terms in \eqref{eq:BBBdBn2} should be present. In particular, we learn that $\EuScript{K}=0$ and $b_0-\wt{b}_0$ should annihilate onshell or offshell vertex operators~$\h{\A}_a$ on which this measure contribution acts, namely
$
\EuScript{R}_{\tilde{z}\theta}=\bm{\theta}D_\theta\EuScript{R}_{\tilde{z}\theta}$ and $\big(b_0-\wt{b}_0\big)\h{\A}_a=0$.
The fact that \smash{$\big(b_0-\wt{b}_0\big)$} appears with coefficient $D_\theta\EuScript{R}_{\tilde{z}\theta}$ in \eqref{eq:BBBdBn2} indicates that the requirement~\smash{$\big(b_0-\wt{b}_0\big)\h{\A}_a=0$} has global origins, whereas the former, since $D_\theta\EuScript{R}_{\tilde{z}\theta}=2$, provides the precise relation between super curvature, $\EuScript{R}_{\tilde{z}\theta}$, and the odd modulus, $\bm{\theta}$. Notice also that ${D_{\bm{\theta}}\EuScript{R}_{\tilde{z}\theta}=D_\theta\EuScript{R}_{\tilde{z}\theta}}$. (The analogous expression in terms of the $\wt{v};\!v|\chi$ supermodulus was given in \eqref{eq:supercurv2}.)
All in all, the final result for the BRST commutator associated to the measure contribution that generates smooth translations of NS punctures across the super Riemann surface~is
\begin{gather*}
\big\{Q_B,\h{\EuScript{B}}_{\tilde{\bm{z}}}\h{\EuScript{B}}_{\bm{z}}\delta\big(\h{\EuScript{B}}_{\bm{\theta}}+\bm{\theta}\h{\EuScript{B}}_{\bm{z}}\big)\big]
= -D_\theta\EuScript{R}_{\tilde{z}\theta}\big(b_0-\wt{b}_0\big)\delta\big(\h{\EuScript{B}}_{\bm{\theta}}+\bm{\theta}\h{\EuScript{B}}_{\bm{z}}\big)+ \h{\partial}_{\tilde{\bm{z}}}^{\rm total}\big(\h{\EuScript{B}}_{\bm{z}}\delta\big(\h{\EuScript{B}}_{\bm{\theta}}+\bm{\theta}\h{\EuScript{B}}_{\bm{z}}\big)\big)\\
\hphantom{\big\{Q_B,\h{\EuScript{B}}_{\tilde{\bm{z}}}\h{\EuScript{B}}_{\bm{z}}\delta\big(\h{\EuScript{B}}_{\bm{\theta}}+\bm{\theta}\h{\EuScript{B}}_{\bm{z}}\big)\big]=}{}
- \h{\partial}_{\bm{z}}^{\rm total}\big(\h{\EuScript{B}}_{\tilde{\bm{z}}}\delta\big(\h{\EuScript{B}}_{\bm{\theta}}+\bm{\theta}\h{\EuScript{B}}_{\bm{z}}\big)\big)
+\h{\partial}_{\bm{\theta}}^{\rm total}\big(\h{\EuScript{B}}_{\tilde{\bm{z}}}\h{\EuScript{B}}_{\bm{z}}\delta'\big(\h{\EuScript{B}}_{\bm{\theta}}+\bm{\theta}\h{\EuScript{B}}_{\bm{z}}\big)\big).
\end{gather*}
In fact, taking into account the relations \eqref{eq:partialIt} and \eqref{eq:hpartialtot}, we arrive at the following result if we also wish to explicitly include the contribution of the worldsheet action~$I$ in the path integral
\begin{gather}
{\rm e}^{-I}\big\{Q_B,\h{\EuScript{B}}_{\tilde{\bm{z}}}\h{\EuScript{B}}_{\bm{z}}\delta\big(\h{\EuScript{B}}_{\bm{\theta}}+\bm{\theta}\h{\EuScript{B}}_{\bm{z}}\big)\big]\nonumber\\
\qquad=-{\rm e}^{-I}D_\theta\EuScript{R}_{\tilde{z}\theta}\big(b_0-\wt{b}_0\big)\delta\big(\h{\EuScript{B}}_{\bm{\theta}}+\bm{\theta}\h{\EuScript{B}}_{\bm{z}}\big)
+ \partial_{\tilde{\bm{z}}}\big({\rm e}^{-I}\h{\EuScript{B}}_{\bm{z}}\delta\big(\h{\EuScript{B}}_{\bm{\theta}}+\bm{\theta}\h{\EuScript{B}}_{\bm{z}}\big)\big)
\nonumber\\
\phantom{\qquad=}{}- \partial_{\bm{z}}\big({\rm e}^{-I}\h{\EuScript{B}}_{\tilde{\bm{z}}}\delta\big(\h{\EuScript{B}}_{\bm{\theta}}+\bm{\theta}\h{\EuScript{B}}_{\bm{z}}\big)\big)+\partial_{\bm{\theta}}\big({\rm e}^{-I}\h{\EuScript{B}}_{\tilde{\bm{z}}}\h{\EuScript{B}}_{\bm{z}}\delta'\big(\h{\EuScript{B}}_{\bm{\theta}}+\bm{\theta}\h{\EuScript{B}}_{\bm{z}}\big)\big),\label{eq:QBBdeltaBIt}
\end{gather}
so that the derivatives appearing now are just ordinary derivatives, making it entirely manifest that the corresponding contribution to the path integral associated to the insertion of the contribution $\h{\EuScript{B}}_{\tilde{\bm{z}}}\h{\EuScript{B}}_{\bm{z}}\delta\big(\h{\EuScript{B}}_{\bm{\theta}}+\bm{\theta}\h{\EuScript{B}}_{\bm{z}}\big)$ to the measure (as we try to unwrap a BRST contour off the surface to establish the decoupling of BRST-exact states) is a total derivative in supermoduli space. (Incidentally, the insertion $\h{\EuScript{B}}_{\tilde{\bm{z}}}\h{\EuScript{B}}_{\bm{z}}\delta\big(\h{\EuScript{B}}_{\bm{\theta}}+\bm{\theta}\h{\EuScript{B}}_{\bm{z}}\big)$ does not depend on the remaining supermoduli.)

\section{Discussion}\label{sec:D}
The path integral expression for the measure arrived at in \eqref{eq:BBd(B)} or \eqref{eq:BBB2x}, and the corresponding result for the BRST (anti-)commutator \eqref{eq:QBBdeltaBIt} are the main results of the present paper. The relation \eqref{eq:BBd(B)} provides the explicit expression for the path integral measure in heterotic string theory that translates fixed picture NS vertex operators (in the natural $-1$ picture) on a super sphere to integrated picture (corresponding to 0 picture). A crucial property is that the dependence on $\wt{v};\!v|\chi$ is smooth, with super curvature encoded locally (as opposed to in transition functions on patch overlaps \cite{Nelson89}). The BRST anti-commutator \eqref{eq:QBBdeltaBIt} demonstrates that BRST-exact vertex operators decouple from amplitudes up to boundary terms (that come from the ``physical'' boundary of supermoduli space as opposed to fictitious boundaries associated to patch overlaps).\looseness=-1

If there is a second NS vertex operator then we can use the same underlying chart with coordinate $\wt{w};\!w|\psi$ and simply place the first and second vertex operator at, say, $\wt{w};\!w|\psi=\wt{v}^1;\!v^1|\chi^1$ and $\wt{w};\!w|\psi=\wt{v}^2;\!v^2|\chi^2$, respectively, the measure contribution being a product of terms as in~\eqref{eq:BBd(B)} with the obvious replacements. The generalisation to any number of insertions is, similarly, immediate.

It is worth noting that (for the super sphere) only a single coordinate chart $\wt{w};\!w|\psi$ is really needed in this viewpoint in a practical computation. Since super curvature is localised in the bulk of the super sphere, the point at ``infinity'' is ``trivialised'' (it does not contribute, e.g., to the Euler characteristic). (One can of course simply map to the $u$-chart using \eqref{eq:ztheta-S2-trans} to include the missing point at infinity when desirable.) So one may ask to what extent this really addresses the main issue associated to a smooth splitting, since we have not needed multiple coordinate patches to cover the supermoduli space associated to a puncture insertion on the super sphere. The main point is that we have used the invariance under superconformal transformations to pick a specific globally-defined gauge slice in the integral over supermoduli. We have treated even and odd moduli on equal footing. The dependence of the resulting measure on these supermoduli is smooth, so that one can use smooth transition functions to transition to a different patch. The only remaining symmetry is a $\mathrm{SU}(1)$ residual symmetry, corresponding to a phase that cannot be fixed globally due to the non-vanishing Euler number of a super sphere. (It is necessary to check that amplitudes do not depend on this phase.)

We have also demonstrated (in an explicit example but the prescription is generally applicable) how it is possible to have a smooth dependence of the superconformal transition functions (defining a super Riemann surface) on supermoduli, while retaining our standard superconformal field theory techniques that have a well-defined notion of chiral and anti-chiral halves. (As we briefly skimmed through in the introduction, in the standard descriptions it is vital that we are still able to distinguish left- from right-moving degrees of freedom in order to even define the theory, and this is all perfectly consistent with the smooth gauge slice we have constructed.)

In particular, we are still defining a super Riemann surface at fixed complex structure using superconformal transition functions, out of which corresponding superconformal frames can be constructed. And these, in turn, can be adopted to construct mode expansions, states, and local NS vertex operators, etc., just as one is used to doing in a corresponding radial quantisation. For this reason, it is also clear how to sum over spin structures for left- and right-moving modes independently. The measure contributions that we have derived here, in a sense, translate all of that to ``integrated picture'', incorporating local super curvature as necessary. What is happening, essentially, is that there is a well-defined notion of a chiral or anti-chiral half in the fixed (or $-1$) picture vertex operators, whereas when we go to integrated (or 0) picture this distinction becomes somewhat obscured due to the presence of super curvature. (Incidentally, since it only involves superghost contributions, the worldsheet path integral measure we have derived is also independent of the string background.)

Elaborating a little, fixed-picture vertex operators \smash{$\h{\A}_a^{(z)}$}, where $a$ labels the state, on which the operator \smash{$\h{\EuScript{B}}_{\tilde{v}}\h{\EuScript{B}}_{v}\delta\big(\h{\EuScript{B}}_{\chi}\big)$} acts can be constructed using any one of the familiar techniques, such as radial quantisation. (The frame label $^{(z)}$ can be identified with the `superconformal normal coordinates' that we have constructed.) So, in particular, a fixed-picture vertex operator will have a well-defined notion of a chiral or anti-chiral half. It has not been at all obvious in the past that this is possible, that it is possible to have a clear distinction between chiral and anti-chiral halves while still having implemented a smooth gauge slice in the integral over supermoduli. Pursuing this further, it is perhaps useful to note that the full set of offshell fixed (or $-$1) picture vertex operators can be derived by cutting open the path integral across, say, an $A_I$-cycle. For the states in the NS sector, this is effectively implemented by inserting into the path integral a resolution of unity \cite{LaNelson90}
\begin{gather}
\suminnt\limits_{a}    \hat{\A}_a^{(z_1)}|0\rangle^1\otimes\hat{\A}^a_{(z_2/q)}|0\rangle^2
= \frac{\alpha'g_D^2}{8\pi {\rm i}}\int \frac{\rmd^Dk}{(2\pi)^D}{\rm e}^{{\rm i}k\cdot (x_0^{(z_1)}-x_0^{(z_2)})}\bar{q}^{\frac{\alpha'}{4}k^2-1}q^{\frac{\alpha'}{4}k^2-1/2}\nonumber\\
\phantom{\qquad}{}\times \exp\left[\sum_{n=1}^{\infty}\bar{q}^n\left(-\frac{1}{n}\tilde{\alpha}_{-n}^{(z_1)}\cdot \tilde{\alpha}_{-n}^{(z_2)}+\tilde{c}_{-n}^{(z_1)}\tilde{b}_{-n}^{(z_2)}-\tilde{b}_{-n}^{(z_1)}\tilde{c}_{-n}^{(z_2)}\right)\right]\nonumber\\
\phantom{\qquad}{}\times\exp \left[\sum_{n=1}^{\infty}i\eta \bar{q}^{ n-1/2}\big(\lambda^{(z_1)}_{-n+1/2}\cdot \lambda^{(z_2)}_{-n+1/2}\big)\right]\nonumber\\
\phantom{\qquad}{}\times\exp \left[\sum_{n=1}^{\infty}q^n\left(-\frac{1}{n}\alpha_{-n}^{(z_1)}\cdot \alpha_{-n}^{(z_2)}+c_{-n}^{(z_1)}b_{-n}^{(z_2)}-b_{-n}^{(z_1)}c_{-n}^{(z_2)}\right)\right]\nonumber\\
\phantom{\qquad}{}\times\exp \left[\sum_{n=1}^{\infty}{\rm i}\eta q^{n-1/2}\big(\beta_{-n+1/2}^{(z_1)} \gamma_{-n+1/2}^{(z_2)}-\gamma_{-n+1/2}^{(z_1)} \beta_{-n+1/2}^{(z_2)}+\psi^{(z_1)}_{-n+1/2}\cdot\psi^{(z_2)}_{-n+1/2}\big)\right]\nonumber\\
\phantom{\qquad}{}\times\eta\big[1+\big(\tilde{c}_0^{(z_1)}+\tilde{c}_0^{(z_2)}\big)\big(c_0^{(z_1)}+c_0^{(z_2)}\big)\big]\nonumber\\
\phantom{\qquad}{}\times
\tilde{c}_1^{(z_1)}c_1^{(z_1)}\tilde{c}_1^{(z_2)}c_1^{(z_2)}\delta\big(\gamma_{1/2}^{(z_1)}\big)\delta\big(\gamma_{1/2}^{(z_2)}\big)|0\rangle^{1}\otimes|0\rangle^{2},\label{eq:handle}
\end{gather}
where the various modes have been defined in Appendix \ref{sec:ME}, whereas $D$ denotes the number of non-compact spacetime dimensions $D=10$. (For $D<10$ one needs to include some additional states in the resolution of unity depending on the compactification manifold, including a discrete sum over any corresponding momentum and winding modes if the compactification allows for these, etc.) The frames $z_1|\theta_1$ and $z_2|\theta_2$ are glued on an annular patch overlap $\U_1\cap \U_2$ with the resulting transition functions $z_1z_2=-\varepsilon^2$ subject to $D_{\theta_2}z_1=\theta_1D_{\theta_2}\theta_1$,
with $q=-\varepsilon^2$, or, more explicitly,\looseness=-1
\begin{gather*}
z_1(z_2|\theta_2) = \frac{-\varepsilon^2}{z_2},\qquad
\theta_1(z_2|\theta_2)=\eta\varepsilon\frac{\theta_2}{z_2}
\qquad {\rm with}\quad \eta=\pm1,
\end{gather*}
with a similar relation for the anti-chiral half, $\wt{z}_1\wt{z}_2=\wt{q}$.
It is important to realise that we do not need to glue with a more general transition function, such as $(z_1-\theta_1\vartheta_1)(z_2-\theta_2\vartheta_2)=-\varepsilon^2$, because the map to integrated picture already incorporates the effect of the odd moduli (analogous to~$\vartheta_1$,~$\vartheta_2$). (The map to integrated picture also captures the effects of super curvature, which become important when the handle moves across the underlying curved surface.) The sign, $\eta$, (unlike for the sphere) here plays an important role: the two values define the two NS spin structures: summing over it in the path integral gives the GSO projection in the NS sector (see~\cite[Section~6.2.3]{Witten12b}). We also took into account that \[
\delta\big(\gamma_{1/2}^{(z_2/q)}\big) = q^{1/2}\delta\big(\gamma_{1/2}^{(z_2)}\big).
\]

Any one particular fixed picture (or $-1$ picture) offshell vertex operator \smash{$\h{\A}_a^{(z)}$} can be derived from this expression \eqref{eq:handle} by expanding it in powers of the pinch parameters $q=-\varepsilon^2$, $\bar{q}$, and then identifying the corresponding momentum integrand with the tensor product of the corresponding vertex operator, \smash{$\h{\A}_a^{(z_1)}\h{\A}^a_{(z_2/q)}$} as indicated on the left-hand side in \eqref{eq:handle}. The ``integrated picture'' (or 0-picture) vertex operators are then given by
\begin{align}
\A_a^{(z)} &= \int \mathcal{D}(\wt{v},\!v|\chi){\rm e}^{-I_{\tilde{v}v\chi}}\h{\EuScript{B}}_{\tilde{v}}\h{\EuScript{B}}_{v}\delta\big(\h{\EuScript{B}}_{\chi}\big)\h{\A}_a^{(z)}\nonumber\\
&=\int \mathcal{D}(\wt{\bm{z}},\!\bm{z}|\bm{\theta}){\rm e}^{-I_{\tilde{\bm{z}}\bm{z}\bm{\theta}}}
\h{\EuScript{B}}_{\tilde{\bm{z}}}\h{\EuScript{B}}_{\bm{z}}\delta\big(\h{\EuScript{B}}_{\bm{\theta}}+\bm{\theta}\h{\EuScript{B}}_{\bm{z}}\big)\h{\A}_a^{(z)},\label{eq:0picvert}
\end{align}
with the measure given in \eqref{eq:BBd(B)} or \eqref{eq:BBB2x}. (To actually carry out this integral it is useful to rewrite the vertex operator integrands in \eqref{eq:0picvert} in terms of the $^{(w)}$ frame coordinates, because this will expose any remaining implicit $\wt{v},\!v|\chi$ dependence while placing all vertex operators in the same frame. Ultimately, in order to perform the relevant Wick contractions when we evaluate the path integral, it should be easiest if the composite operators whose contractions we are computing are defined in the same frame, the natural choice being the $^{(w)}$ frame (despite the fact that they will usually be inserted at different coordinate values). That is, after extracting the integrated (0 picture) vertex operators of interest, it may be desirable to map the supermoduli integrands back to the $^{(w)}$ chart coordinates.

Actually, this cutting open of the path integral is appropriate when we treat all $A_I$-cycle loops on equal footing (i.e., we need to cut open all loops and insert a resolution of unity in each one), because it is only in this case that a super Riemann surface with an arbitrary number of loops can be mapped to a super Riemann surface with the topology of a 2-sphere, although this can be relaxed depending on the objective.

Notice that, as we have already mentioned, there is a clear distinction between chiral and anti-chiral halves in \smash{$\h{\A}_a^{(z)}$}, and that it is the map of non-primary vertex operators to integrated picture~\smash{$\A_a^{(z)}$} that obscures the distinction between chiral and anti-chiral halves. This observation is important, because it resolves the question of how to sum over spin structures in the presence of a smooth gauge slice. Actually, the resolution is reminiscent of the D'Hoker and Phong resolution~\cite{D'HokerPhong89}, which showed that at fixed internal loop momenta there is a useful notion of chiral splitting, which in turn made it clear how to sum over spin structures and hence incorporate the GSO projection \cite{SeibergWitten86}. Although the original proof of chiral splitting does not hold at arbitrarily high genus due to the Donagi--Witten obstruction \cite{DonagiWitten15}, the approach that we are presenting here does (modulo Ramond sector contributions that have not yet been worked out). The correspondence with the D'Hoker and Phong procedure \cite{D'HokerPhong89} becomes apparent when one takes into account that the present approach fixes not only the internal loop momenta, but instead fixes \emph{all} quantum numbers that characterise the offshell state of a string propagating through an internal loop (including the loop momenta). The resolution of unity \eqref{eq:handle} makes this fully explicit. This is quite natural, in that if the formalism simplifies when we fix the internal loop momenta \cite{D'HokerPhong89}, it might be sensible to try to also fix the remaining quantum numbers of these internal string states. This turns out to be a good idea, because it generalises immediately to an arbitrary number of string loops while circumventing the obstructions~\cite{DonagiWitten15} that appeared in the original formalism~\cite{D'HokerPhong89}. There is some related discussion in the context of bosonic string theory in~\cite{SklirosLuest21}.

Let us also elaborate on the situation when the underlying super Riemann surface has arbitrary local super curvature.
One of the key ingredients in constructing the smooth gauge slice that associates a supermodulus of dimension $2|1$ to a NS puncture are the superconformal vector fields given in \eqref{eq:deltaVtdeltaVdztdzdthbf} that are associated to an underlying super Riemann surface with the topology of a 2-sphere and with super curvature, loosely speaking, evenly distributed throughout the surface. A natural question is to understand how these superconformal vector fields would change if we instead considered arbitrary local super curvature (which might correspond to a super Riemann surface with any number of handles, rather than a super Riemann surface whose reduced space is a 2-sphere with constant curvature). This will be discussed elsewhere, but I wish to remark that this situation \emph{is} more subtle and not entirely clear. With regards to gauge invariance, on a general super Riemann surface with an arbitrary number of handles and arbitrary local super curvature, there is a result similar to \eqref{eq:QBBdeltaBIt}, but it is not too hard to show that there also appear certain additive terms on the right-hand side that depend on bilinear products of $\wt{\bm{z}}$, $\bm{z}$ derivatives of the super curvature evaluated at the puncture. So if the super curvature $\R_{\tilde{z}\theta}$ is proportional to the odd modulus $\bm{\theta}$ of the NS puncture when evaluated at the NS puncture in question, $\wt{z};\!z|\theta=0;\!0|0$, these additional squared super curvature terms vanish and gauge invariance is restored. However, it is not obvious if a more general dependence of the super curvature on the odd modulus is allowed. In general, we also expect the BRST charge to receive corrections at higher genus. The case of arbitrary super curvature and arbitrary-genus super Riemann surfaces certainly deserves further study. But it should be kept in mind that in the handle operator viewpoint \cite{SklirosLuest21}, where all genus amplitudes are constructed on the 2-sphere with additional handle operator insertions (by an appropriate cutting and gluing procedure), the case of the super sphere discussed in this article should actually suffice.

Incidentally, for the reader that is wondering whether handle operator insertions on a 2-sphere can really capture the full supermoduli space of super Riemann surfaces, one can outline a~procedure to demonstrate this explicitly (by cutting and gluing the path integral across different cycles, and then changing variables in the integral over supermoduli). An illustrative case is to start from $\g$ handle operators on the 2-sphere, consider any two such handle operators, and then cut open the path integral across a closed curve on the 2-sphere that contains one of the two handle operators entirely, and also contains one of the two local insertions associated to the second handle operator. Insert a complete set of states on the corresponding cut, and then change variables in the integral over supermoduli so as to expose the modulus that gives rise to the propagator in the cut cycle. The resulting configuration is that of a double handle (one handle inserted on another handle). This illustrates the point (albeit with an example) that there are regions in supermoduli space that are already included in the original ($\g$ handle operators on a 2-sphere) setup that one might have naively thought were not included. More generally, we already know that this must have been the case due to \cite{Polchinski_v1,Polchinski_v2} OPE associativity, modular invariance, and invariance of the path integral measure under super reparametrisations in supermoduli space. (One can try to prove that the resulting amplitudes are modular invariant, since this has not been shown in full generality. Some initial discussion along these lines at one loop and for the bosonic string can be found in \cite{SklirosLuest21}, and although I do not expect any surprises at higher loop orders, it would be satisfying to see modular invariance demonstrated more explicitly in the handle operator viewpoint at arbitrary loop order.)

There is clearly a lot a work that remains to be done. It would also be interesting, as a warmup, to calculate the dilaton one-point amplitude (or to derive the dilaton theorem) \cite{BergmanZwiebach95,DistlerNelson91,Erler22,LaNelson90,Polchinski87,Polchinski88} in this formalism. The reason being that the dilaton is not a primary operator, so that the additional terms derived above that appear due to super curvature are expected to contribute non-trivially. Some recent discussion of the dilaton theorem that makes use of Polchinski's original bosonic string gauge slice \cite{Polchinski88} (the heterotic string generalisation of which was developed in the present document) can be found in \cite{SenZwiebach24} (which among other things summarises \cite{BergmanZwiebach95}). There is some related work in the appendix of \cite{Polchinski87} where the same effect arises by storing local curvature in the transitions functions instead. One might envision an analogous heterotic string theory computation making use of the gauge slice developed in the current document. A more ambitious direction is to construct the full handle operators associated to this gauge slice. One reason being that inspired by an idea in \cite{Tseytlin90} (which inspired the detailed study in \cite{SklirosLuest21}), one can then ask under what conditions we might be able to sum over handle operators (which corresponds to summing over string loops at the level of the integrand). Since all loop orders are treated on equal footing, i.e., one handle operator insertion for every string loop inserted on a super sphere, it is tempting to speculate whether one might even be able to go beyond perturbation theory in this manner (see also the related discussion in the introduction); and if so, under what assumptions.

The first step however is presumably to unravel how to implement this smooth gauge slice in the Ramond sector, because, needless to say, it is clearly necessary \cite{SeibergWitten86} to include both NS and R sectors in any handle operator that is meant to exactly incorporate the full implications of a~string loop insertion on a super sphere.

It is worth emphasising that despite efforts in \cite{SklirosLuest21} and in the current document, the dif\-fer\-en\-tial-geometric (or handle operator) approach to the worldsheet approach to (super)string theory is of course much less developed than the corresponding (super)string field theory approach. It is well known that in a generic amplitude there are infrared divergences \cite{Witten13b} that require some kind of analytic continuation (when the internal momenta are sufficiently generic), and sometimes such analytic continuation is not even possible, in which case it has been shown that one can resolve any difficulties encountered by appealing to (super)string field theory, see~\cite{SenZwiebach24} for a recent discussion and references therein. More generally, there are perturbative states that require mass and wavefunction renormalisation, and there are also tadpole divergences that one needs to understand how to cancel. One expects that if the handle operator approach can also resolve these difficulties, they will presumably be resolved in a way that is fundamentally distinct from the corresponding (super)string field theory resolutions (because the supermoduli space parametrisation is fundamentally different), but it is still too early to elaborate on these issues.\looseness=-1

A related future direction is to understand how to gauge fix the invariance under $\mathrm{OSp}(2,1)$ in a manner that does not depend on the number of vertex operator insertions. Once we have reduced a general genus-$\g$ amplitude to a sphere with handle operator insertions, we also recover the underlying symmetries of the super sphere, which in turn need to be fixed (see~\cite{AhmadainWall22,AnninosBautistaMuhlmann21,Eberhardt20,EberhardtPal21,ErbinMaldacenaSkliros19,Erler22,MahajanStanfordYan21}). Furthermore, if would be interesting to understand how modular invariance is restored when we try to sum over string loops. What role might the background symmetries play in this story? At a more basic level, the formalism presented here immediately applies to offshell string theory in the BRST formalism (because we have constructed a globally-defined gauge slice), so one can use it to ask various questions where going offshell is important, see e.g., \cite{AhmadainWall22,AhmadainWall22b,Sen15b}, and in particular when non-primary vertex operators contribute (which is why we mentioned the dilaton theorem above).

\appendix

\section{Derivation}\label{sec:MC}

In this appendix, we will compute the derivatives appearing in \eqref{eq:VtVcomp}.
Taking into account \eqref{eq:superconftrans}, and in particular
$z(w|\psi)=\frac{w-v-\psi\chi}{\wt{v}w+1}$,
it follows that the variation $\delta z$ is given by
\begin{align}
\delta z &= \left[\delta\wt{v} \frac{\partial z}{\partial \wt{v}}+\delta v \frac{\partial z}{\partial v}+\delta\chi  \frac{\partial z}{\partial \chi}\right]_{w|\psi}\nonumber\\
&=
\delta\wt{v} \left(\frac{(w-v-\psi\chi)w}{-(\wt{v}w+1)^2}\right)
+\delta v\left(\frac{-1}{\wt{v}w+1}\right)
+\delta\chi \left(\frac{\psi}{\wt{v}w+1}\right),\label{eq:deltaz}
\end{align}
where the derivatives with respect to the supermoduli, $\wt{v};\!v|\chi$, are evaluated at fixed $\wt{w};\!w|\psi$. As seen in \eqref{eq:Bthat3}, for the path integral measure we actually need these variations in terms of $\wt{z};\!z|\theta$ rather than $\wt{w};\!w|\psi$. We eliminate the dependence on the latter in favour of the former by making use of the inverse relations \eqref{eq:wtwpsiOSp},
\begin{gather*}
w(z|\theta) = \frac{z+v+\eta\theta\chi/\sqrt{1+\wt{v}v}}{-\wt{v}z+1-\eta\theta\chi \wt{v}/\sqrt{1+\wt{v}v}},\qquad
\psi(z|\theta)= \frac{\sqrt{1+\wt{v}v} \eta\theta+\chi}{-\wt{v}z+1}.
\end{gather*}
Before explicitly displaying the resulting expression for $\delta z$ however, we recall that there is also some information about the phase of $z+\delta z(z|\theta)$ in \eqref{eq:deltaz}, which as we have discussed does not have physical significance. So we would like to extract this, not because it is necessary, but because it is simplest to set it equal to any convenient value, and then check that quantities of physical interest do not depend on that choice. Adding $z$ to both sides of \eqref{eq:deltaz} and taking the preceding comments into account, to leading order in the variation, we see that
\begin{align*}
z+\delta z(z|\theta) ={}& {\rm e}^{2{\rm i} \operatorname{Im} \frac{\tilde{v}\delta v}{1+\tilde{v}v}}\left\{z -\frac{\delta\wt{v}}{1+\wt{v}v} \left(z^2-\frac{2\eta\chi}{\sqrt{1+\wt{v}v}}\frac{1}{2}z\theta\right)
-\frac{\delta v}{1+\wt{v}v}\left(1+\frac{2\eta\chi}{\sqrt{1+\wt{v}v}}\frac{1}{2}\theta\wt{v}\right)\right.\\
& \left.+\frac{\eta\delta\chi}{\sqrt{1+\wt{v}v}} \left(\theta+\frac{\eta\chi}{\sqrt{1+\wt{v}v}}\right)\right\},
\end{align*}
so that we can now easily identify the overall phase. Dropping this, we learn that the variation takes the form
\begin{align*}
\delta z(z|\theta) ={}& -\frac{\delta\wt{v}}{1+\wt{v}v} \left(z^2-\frac{2\eta\chi}{\sqrt{1+\wt{v}v}}\frac{1}{2}z\theta\right)
-\frac{\delta v}{1+\wt{v}v}\left(1+\frac{2\eta\chi}{\sqrt{1+\wt{v}v}}\frac{1}{2}\wt{v}\theta\right)\\
&+\frac{\eta\delta\chi}{\sqrt{1+\wt{v}v}} \left(\frac{\eta\chi}{\sqrt{1+\wt{v}v}}+\theta\right).
\end{align*}

The corresponding variation $\delta\theta$ at generic $z|\theta$ is similarly determined from \eqref{eq:superconftrans}, namely
\begin{gather*}
\theta(w|\psi)
=\eta \frac{\sqrt{1+\wt{v}v}}{(\wt{v}w+1)}\psi-\eta \frac{\chi}{\sqrt{1+\wt{v}v}},\qquad\eta=\pm1,
\end{gather*}
and found to be given by
\begin{align*}
\delta\theta(z|\theta)
&= \left[\delta\wt{v} \frac{\partial \theta}{\partial \wt{v}}\theta+\delta v \frac{\partial \theta}{\partial v}\theta+\delta\chi  \frac{\partial \theta}{\partial \chi}\theta\right]_{w|\psi}\\
&= -\frac{\delta\wt{v}}{1+v\wt{v}}
\left(z\theta+\frac{1}{2}v\theta
+\frac{z\eta \chi}{\sqrt{1+v\wt{v}}}
\right)+
\frac{\delta v}{1+v\wt{v}} \left(\frac{1}{2}\wt{v}\theta+\frac{\wt{v}\eta\chi}{\sqrt{1+v\wt{v}}}\right)-\frac{\eta \delta\chi}{\sqrt{1+v\wt{v}}}.
\end{align*}
Extracting the overall phase as above, and adding this variation to $\theta$, we find that to leading order in the variations
\[
\theta+\delta\theta(z|\theta)={\rm e}^{{\rm i}\operatorname{Im}\frac{\tilde{v}\delta v}{1+v\tilde{v}}}\left\{\theta
-\frac{\delta\wt{v}}{1+v\wt{v}}
\left(z\theta
+\frac{z\eta \chi}{\sqrt{1+v\wt{v}}}
\right)+
\frac{\delta v}{1+v\wt{v}} \left(\frac{\wt{v}\eta\chi}{\sqrt{1+v\wt{v}}}\right)-\frac{\eta \delta\chi}{\sqrt{1+v\wt{v}}}\right\}.
\]
Dropping the overall phase (which is half that found in $\delta z$ as required by the superconformal condition), we learn that
\begin{gather*}
\delta \theta(z|\theta) = -\frac{\delta\wt{v}}{1+v\wt{v}}
\left(z\theta
+\frac{\eta \chi}{\sqrt{1+v\wt{v}}}z
\right)+
\frac{\delta v}{1+v\wt{v}} \left(\frac{\eta\chi }{\sqrt{1+v\wt{v}}}\wt{v}\right)-\frac{\eta \delta\chi}{\sqrt{1+v\wt{v}}}.
\end{gather*}
The quantity that appears in the path integral measure are actually components of the superfield $\delta\V\equiv \delta z-\delta\theta \theta$ corresponding globally to a section of \mbox{$T\Sigma/\mathcal{D}\cong\mathcal{D}^2$}
\begin{align}
\delta\V^{(z)} ={}&
-\frac{\delta\wt{v}}{1+\wt{v}v} \left(z^2-\frac{2\eta\chi}{\sqrt{1+\wt{v}v}}z\theta\right)
-\frac{\delta v}{1+\wt{v}v}\left(
1+\frac{ 2\eta\chi }{\sqrt{1+\wt{v}v}}\wt{v}\theta\right)\nonumber\\
&+\frac{\eta\delta\chi}{\sqrt{1+\wt{v}v}} \left(2\theta+\frac{\eta\chi}{\sqrt{1+\wt{v}v}}\right).\label{eq:deltaVn}
\end{align}

An analogous computation for the anti-chiral half $\delta\wt{z}(\wt{w})|_{\wt{w} \textrm{fixed}}=\delta\wt{v} \partial_{\tilde{v}}\wt{z}(\wt{w})+\delta v \partial_v\wt{z}(\wt{w})+\delta\chi \partial_{\chi}\wt{z}(\wt{w})$ according to \eqref{eq:superconftrans} and \eqref{eq:wtwpsiOSp} and by a slight abuse of notation denoting this again by $\delta \wt{z}(\wt{z})$ yields
\begin{gather*}
\delta \wt{z}(\wt{z})
=-\frac{\delta\wt{v}}{1+\wt{v}v} \big(1-v\wt{z}\big)
-\frac{\delta v}{1+\wt{v}v} \big(\wt{z}^2+\wt{v}\wt{z}\big),
\end{gather*}
so that writing $\delta\wt{\V}^{(z)}(\wt{z}) = \delta \wt{z}(\wt{z})$ and extracting out the phase again and dropping it, we arrive~at
\begin{gather}\label{eq:deltazt}
\delta\wt{\V}^{(z)}(\wt{z})
=-\frac{\delta\wt{v}}{1+\wt{v}v}
-\frac{\delta v}{1+\wt{v}v} \wt{z}^2.
\end{gather}

It will prove efficient to introduce notation for the variations with respect to specific moduli in \eqref{eq:deltazt} and \eqref{eq:deltaVn}. Define
\begin{gather*}
\delta\wt{\V}^{(z)}(\wt{z})= \delta\wt{v}\wt{\V}_{\tilde{v}}(\wt{z})+\delta v\wt{\V}_v(\wt{z})+\delta\chi\wt{\V}_{\chi}(\wt{z}),\\
\delta\V^{(z)}(z|\theta)=\delta\wt{v}\V_{\tilde{v}}(z|\theta)+\delta v\V_v(z|\theta)+\delta\chi\V_{\chi}(z|\theta),
\end{gather*}
so that according to \eqref{eq:deltazt} and \eqref{eq:deltaVn}
\begin{alignat*}{3}
& \wt{\V}_{\tilde{v}}(\wt{z}) =-\frac{1}{1+\wt{v}v},\qquad && \V_{\tilde{v}}(z|\theta) =-\frac{1}{1+\wt{v}v}\left(z^2-\frac{2\eta\chi}{\sqrt{1+\wt{v}v}}z\theta\right),&\\
&\wt{\V}_v(\wt{z})= -\frac{\wt{z}^2}{1+\wt{v}v},\qquad && \V_v(z|\theta)= -\frac{1}{1+\wt{v}v}\left(
1+\frac{ 2\eta\chi \wt{v}}{\sqrt{1+\wt{v}v}}\theta\right),&\\
&\wt{\V}_{\chi}(\wt{z}) =0, \qquad && \V_{\chi} (z|\theta)=\frac{\eta}{\sqrt{1+\wt{v}v}}
\left(2\theta+\frac{\eta\chi}{\sqrt{1+\wt{v}v}}\right).&
\end{alignat*}

\section{Mode expansions}\label{sec:ME}
Let us consider a local frame, $(U,\tilde{z};\!z|\theta)$, and mode expand the various matter and ghost superfields around $\wt{z};\!z|\theta=0;\!0|0$. We will restrict attention to the NS sector. Neglecting auxiliary fields, for the chiral half of the superghosts, in particular, we write
\begin{gather*}
B_{z\theta}=\beta(z)+\theta b(z)=\sum_{n\in\mathbb{Z}}\frac{\beta_{n+1/2}+\theta b_n}{z^{n+2}},\qquad
C^z = c(z)+\theta\gamma(z)= \sum_{n\in\mathbb{Z}}\frac{c_n+\theta\gamma_{n-1/2}}{z^{n-1}},
\end{gather*}
whereas for the anti-chiral halves
\begin{gather*}
\wt{B}_{\tilde{z}\tilde{z}}=\wt{b}(\wt{z})=\sum_{n\in\mathbb{Z}}\frac{\wt{b}_n}{\wt{z}^{n+2}}
\qquad{\rm and}\qquad
\wt{C}^{\tilde{z}}=\wt{c}(\wt{z})=\sum_{n\in\mathbb{Z}}\frac{\wt{c}_n}{\wt{z}^{n-1}}.
\end{gather*}
Similarly, for the chiral half of the matter fields the mode expansions are
\begin{gather*}
D_\theta X^\mu(z|\theta) = \psi^\mu(z)+\theta \partial_z x^\mu(z)=\sum_{n\in\mathbb{Z}}\frac{\psi_{n+1/2}^\mu-{\rm i}\theta\alpha_n^\mu}{z^{n+1}},
\end{gather*}
whereas for the anti-chiral half of the matter fields
\begin{gather*}
\partial_{\tilde{z}} X^\mu(\wt{z}) = \partial_{\tilde{z}} x^\mu(\wt{z})=-{\rm i}\sum_{n\in\mathbb{Z}}\frac{\wt{\alpha}_n^\mu}{z^{n+1}}
\qquad{\rm and}\qquad
\Lambda_a(\tilde{z}) = \lambda_a(\tilde{z})=\sum_{n\in\mathbb{Z}}\frac{\tilde{\lambda}_{n+1/2}^a}{\wt{z}^{n+1}}.
\end{gather*}
We can now define the $\mathrm{OSp}(2,1)$ vacuum denoted by $|0\rangle$, which is analogous to the $\mathrm{SL}(2,\mathbb{C})$ vacuum for ordinary Riemann surfaces. This is simply the state corresponding to the unit operator, so is defined in the NS sector by
\begin{alignat*}{5}
&\beta_{n+1/2}|0\rangle = 0,\qquad &&n\geq-1,\qquad&&
\gamma_{n-1/2}|0\rangle = 0,\qquad&& n\geq2,&\\
&\wt{b}_{n}|1\rangle=b_{n}|0\rangle = 0,\qquad&& n\geq-1,\qquad&&
\wt{c}_{n}|1\rangle=c_{n}|0\rangle = 0,\qquad&& n\geq2,&\\
&\wt{\alpha}_{n}|0\rangle=\alpha_{n}|0\rangle = 0,\qquad&&n\geq0,\qquad&&
\psi_{n+1/2}|0\rangle = 0,\qquad&& n\geq0,&\\
&\lambda_{n+1/2}|0\rangle  =0,\qquad&& n\geq0,&&&&&\\
&\wt{L}_n|0\rangle= L_{n} |0\rangle=G_{n+1/2}|0\rangle = 0,\qquad &&n\geq-1,&&&&&
\end{alignat*}
where it is to be understood that the indices are shifted as necessary such that $n\in\mathbb{Z}$. These are derived by re-expressing the above mode expansions as contour integrals and using that the OPE with the unit operator is non-singular.

The total ghost charge is defined such that $\wt{c}$, $c$, $\gamma$, $\delta(\beta)$ have ghost charge 1, whereas $\wt{b}$, $b$, $\beta$, $\delta(\gamma)$ have ghost charge $-1$.

\subsection*{Acknowledgements}
I am grateful to Eric D'Hoker, Edward Witten and Branislav Jurco for correspondence, to Mark Doyle for sharing his Ph.D.\ Thesis with me, to Imperial College for support, and especially Arkady Tseytlin and the TIFR, Mumbai String Theory group for numerous insightful discussions.

\pdfbookmark[1]{References}{ref}
\LastPageEnding

\end{document}